\documentclass[letter,scriptaddress,aps,noshowkeys,noshowpacs,superscriptaddress]{revtex4}

	\usepackage{natbib}
	\usepackage{amsmath}
	\usepackage{makeidx}
	\usepackage{amsfonts}
	\usepackage[ansinew]{inputenc}
	\usepackage[usenames,dvipsnames]{pstricks}
	\usepackage{subfigure}
	\usepackage{epsfig}
	\usepackage{pst-grad} 
	\usepackage{pst-plot} 
	\usepackage[colorlinks,hyperindex]{hyperref}
	\usepackage{epstopdf}
	\usepackage{multirow}
	\usepackage{titlesec}
	\usepackage{float}
    \usepackage{comment}
    \usepackage[font=small,labelfont=bf]{caption}
    \usepackage{setspace}

	\titlespacing{\section}{0pt}{5pt}{5pt}
	\titlespacing{\subsection}{0pt}{5pt}{5pt}
	\titlespacing{\subsubsection}{0pt}{5pt}{5pt}

	\hypersetup
	{
		colorlinks,%
		citecolor=black,%
		linkcolor=black,%
		urlcolor=black,%
	}

    \newcommand{\smax}{\mbox{\scriptsize{max}}}
    \newcommand{\g}{\mbox{\scriptsize 1D}}
	
\makeindex
\begin{document}

\title{Optical properties of an atomic ensemble coupled to a band edge of a photonic crystal waveguide}

\author{Ewan Munro}
\affiliation{Centre for Quantum Technologies, National University of Singapore, 3 Science Drive 2, 117543 Singapore.}
\author{Leong Chuan Kwek}
\affiliation{Centre for Quantum Technologies, National University of Singapore, 3 Science Drive 2, 117543 Singapore.}
\affiliation{Institute of Advanced Studies, Nanyang Technological University, 60 Nanyang View, Singapore 639673}
\affiliation{National Institute of Education, Nanyang Technological University, 1 Nanyang Walk, Singapore 637616}
\affiliation{MajuLab, CNRS-UNS-NUS-NTU International Joint Research Unit, UMI 3654, Singapore}
\author{Darrick E. Chang}
\affiliation{ICFO-Institut de Ciencies Fotoniques, The Barcelona Institute of Science and Technology, 08860 Castelldefels (Barcelona), Spain}

\date{11 April 2016}
\maketitle

\begin{spacing}{1.6}
We study the optical properties of an ensemble of two-level atoms coupled to a 1D photonic crystal waveguide (PCW), which mediates long-range coherent dipole-dipole interactions between the atoms. We show that the long-range interactions can dramatically alter the linear and nonlinear optical behavior, as compared to a typical atomic ensemble. In particular, in the linear regime, we find that the transmission spectrum reveals multiple transmission dips, whose properties we show how to characterize. In the many-photon regime the system response can be highly non-linear, and under certain circumstances the ensemble can behave like a single two-level system, which is only capable of absorbing and emitting a single excitation at a time. Our results are of direct relevance to atom-PCW experiments that should soon be realizable.
\end{spacing}

\newpage

\section{INTRODUCTION}

Photonic crystal waveguides (PCWs) - dielectric media with a periodically varying refractive index - have attracted significant interest recently as a platform for realizing novel quantum light-matter interfaces. The ability to engineer PCW properties permits control of the electromagnetic environment experienced by nearby atoms, which may be leveraged to achieve strongly enhanced atom-photon coupling efficiencies \cite{thompson2013coupling,gobanAtomLightPCs,LodahlReview,javadi2015single,HouckMar2016} compared to more traditional approaches \cite{ExploringTheQuantum,KimbleQuantumInternet,RempeReviewQuantumNetworks}, as well as for the exploration of new regimes of quantum optics.

\begin{figure}[h]
\centering
\includegraphics[width=0.65\textwidth]{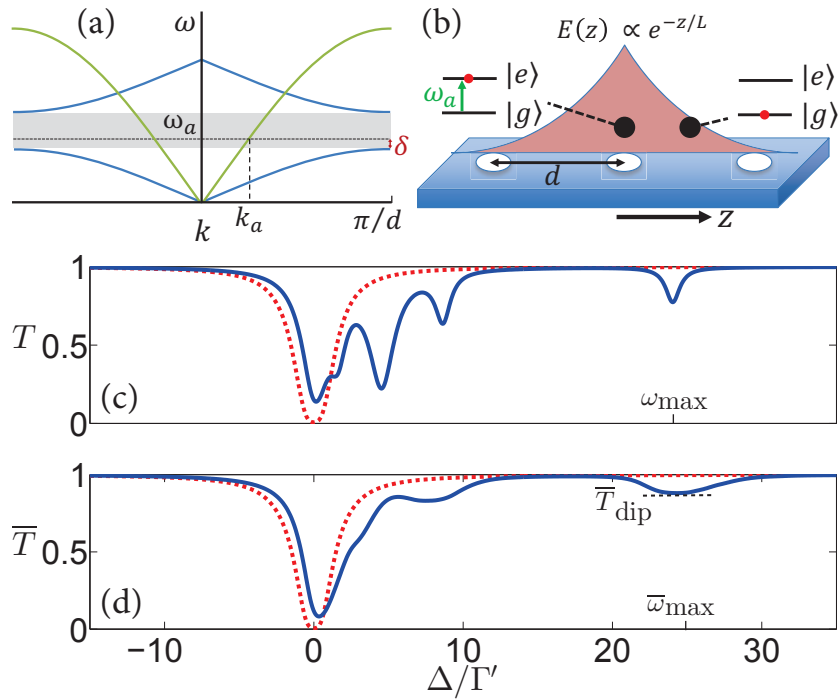}
\caption{(a) Schematic band structure of the PCW, showing the guided-mode frequency $\omega$ vs. Bloch wavevector $k$. The atomic frequency $\omega_a$ is in a band gap (grey region) of one mode (blue), and in a region of linear dispersion of a second mode (green). (b) Illustration of an atom in its excited state $|e\rangle$, coupled to a PCW (in blue). An atom-photon bound state forms when $\omega_a$ is in a band gap: the photonic component of this state (shown in light red) can mediate the exchange of an excitation with a second atom in its ground state $|g\rangle$. (c) Blue solid line: representative single-shot transmission spectrum for $n=10$ atoms placed randomly over $N=200$ sites, with $V/\Gamma^{\prime}=4$, $L/d = 100$, $\Gamma_{\g}/\Gamma^{\prime} = 0.3$ and $k_ad = \pi/2$. Red dashed line: spectrum for same configuration in the absence of long-range interactions ($V/\Gamma^{\prime}=0$), i.e., a normal atomic ensemble. (d) Transmission spectrum for the same parameters as (c), averaged over 1000 samples of atomic positions.}
\label{Fig1}
\end{figure}

An exciting example of the latter \cite{kurizki1990,JohnPhotonHopping,ShamoonNonradiative,ShahmoonNonlocalNonlin,douglas2015quantum,JamesPhotonMolecules,HoodAtomAtominteractions} is the ability to engineer long-range coherent interactions between atoms. The emission properties of an atom coupled to a PCW differ markedly from those in free space when the resonant transition frequency of the atom, $\omega_a$, is inside a band gap, i.e. a frequency domain where guided modes are absent, as depicted in Fig. \ref{Fig1}a for the blue-colored band. Rather than radiating away, the field emitted into this set of modes is localized around the atom by Bragg reflection, giving rise to an atom-photon bound state \cite{JohnWang1990,JohnWang1991,Kofman1994}, shown schematically in Fig. \ref{Fig1}b. The photonic component of this state constitutes an effective cavity mode that can mediate coupling to other atoms, resulting in effective dipole-dipole interactions which acquire the spatial features of the cavity mode. These interactions can in turn be exploited to realize novel regimes of quantum many-body physics and nonlinear optics \cite{ShahmoonNonlocalNonlin,DarrickNatureReview,JamesPhotonMolecules}.

Motivated by remarkable advances to interface atoms or solid-state emitters with PCWs \cite{gobanAtomLightPCs,yu2014nanowire,javadi2015single,HoodAtomAtominteractions}, we study the fundamental optical properties of a system of two-level atoms with long-range interactions, finding rich behavior that differs markedly from an ensemble of independent atoms. For example, as is evident from Fig. \ref{Fig1}c, even the linear transmission spectra can be dramatically modified as compared to a typical atomic ensemble. We will describe the origin of the multiple transmission dips, and show that the resonances associated with some of them exhibit large optical nonlinearities that yield strong photon anti-bunching in the scattered light. Our results are a step towards developing a comprehensive understanding of the linear and nonlinear optical properties of such systems of interacting atoms.

\section{MODEL AND METHODS}

We model the simple case of an ensemble of two-level atoms with ground and excited states $|g\rangle, |e\rangle$, coupled to two guided modes of a 1D PCW, with dispersion relations (frequency $\omega$ vs. Bloch wavevector $k$) as depicted schematically in Fig. \ref{Fig1}a. A realistic platform for implementing such a model is the `alligator' PCW described in Ref. \cite{gobanAtomLightPCs}. The atomic transition frequency, $\omega_a$, is in a band gap of one set of modes, which in practice may have, e.g., transverse electric (TE) polarization. As described above, the absence of propagating states at $\omega_a$ prevents an excited atom from radiating into the TE modes. However, an exponentially decaying photonic cloud of length $L$ can form around the atom, facilitating coherent excitation exchange with a proximal atom in its ground state with a strength $V$. The quantities $V,L$ depend on the curvature of the dispersion relation at the lower band edge and the frequency separation $\delta$ between the band edge and atomic transition, as discussed further in Ref. \cite{douglas2015quantum}. We assume the detuning between $\omega_a$ and the upper TE band edge to be sufficiently large that this band may be neglected.

While the above mechanism can produce novel long-range interactions, it does not allow the atoms to be efficiently probed using the same TE modes, since it is not possible to send in propagating fields at frequency $\omega_a$. To this end, a second band of modes, which may correspond to transverse magnetic (TM) polarization, is used as a conventional waveguide to excite the atoms and collect the resulting scattered fields. Our model is then a minimal description that produces long-range interactions, and allows their effect to be probed. In particular, the level structure is considerably simplified as compared to previous schemes to realize optical nonlinearities in PCWs \cite{ShahmoonNonlocalNonlin,JamesPhotonMolecules}. The coupling strength of a single atom to the TM band is characterized by an emission rate $\Gamma_{\g}$, and the associated wavevector $k_a$ is shown in Fig. \ref{Fig1}a. We assume the atoms are trapped along the axis of the crystal in a lattice of period $d$, equal to the period of the crystal itself. The dynamics of the full system may be described by an effective non-Hermitian Hamiltonian for the atoms alone \cite{chang2012cavity,douglas2015quantum}:

\begin{equation}\label{Htot}
H = -\sum_j^n\left[(\Delta + i\Gamma^{\prime}/2)\sigma_{ee}^j + \Omega(\sigma_{eg}^j e^{ik_L z_j} + h.c.)\right]
+\sum_{j,k}^n \left(V(-1)^{\theta_{jk}}e^{-|z_j - z_k|/L} - i\frac{\Gamma_{\g}}{2}e^{ik_a|z_j - z_k|}\right)\sigma_{eg}^j\sigma_{ge}^k.\\
\end{equation}

Here $\Delta = \omega_L - \omega_a$ denotes the detuning of the probe field (incident from the left with wavevector $k_L$, driving frequency $\omega_L$, and Rabi frequency $\Omega$) from the atomic resonance frequency, $\Gamma^{\prime}$ is the decay rate into free space and all other modes, and $z_j$ is the position of the $j^{th}$ atom along the waveguide. In all simulations we set $\Gamma_{\g}/\Gamma^{\prime}=0.3$, however we will also provide more general conclusions independent of the specific parameter choices. The operator $\sigma_{\mu\nu}^j = |\mu_j\rangle\langle \nu_j|$, where $\mu,\nu = g,e$ are energy eigenstates of the atom $j$. We assume that the atoms are trapped at the anti-nodes, where the interaction with the TE modes is maximized, and results in an integer phase factor $\theta_{jk}=(z_j + z_k)/d$ (as $z_j$ is an integer multiple of $d$).

To obtain the fields scattered into the waveguide, we use the generalized input-output methods described in Ref. \cite{caneva2015quantum}. To preserve the norm of the wave function, dynamics under the non-Hermitian Hamiltonian of Eq. (1) must in principle be supplemented with quantum jumps; however, we will primarily be interested in field correlations under weak driving ($\Omega/\Gamma^{\prime}\ll1$), in which case it can be shown that quantum jumps may be neglected. We time-evolve the initial atomic wavevector $|g\rangle^{\otimes n}$ under Eq. (\ref{Htot}); the transmitted (T) and reflected (R) fields are then reconstructed using the input-output relations

\begin{eqnarray}
a_{\mbox{\scriptsize out,T}}(z) &=& \Omega e^{ik_Lz} + \frac{i\Gamma_{\g}}{2}\sum_j\sigma_{ge}^j e^{ik_a(z-z_j)} + F_{\mbox{\scriptsize T}}, \label{inoutT}\\
a_{\mbox{\scriptsize out,R}}(z) &=& \frac{i\Gamma_{\g}}{2}\sum_j\sigma_{ge}^j e^{-ik_a(z-z_j)} + F_{\mbox{\scriptsize R}},\label{inoutR}
\end{eqnarray}

where the transmitted (reflected) field is evaluated at a position $z$ beyond the last (first) atom of the array (henceforth we drop the subscript \emph{out} and the argument $z$). Physically, Eqs. (\ref{inoutT})-(\ref{inoutR}) state that the output field properties are completely encoded in the input field and the correlations of the atomic scatterers. The vacuum input noise operators $F_{\mbox{\scriptsize T,R}}$ may be neglected on our correlations of interest. The steady-state transmittance and two-photon correlation function of the transmitted field are given by

\begin{eqnarray}
T &=& \frac{\langle\psi|a^{\dagger}_T a_T |\psi\rangle}{\Omega^2},\label{Transmittance}\\
g^{(2)}_T(\tau) &=& \frac{\langle\psi|a^{\dagger}_Te^{iH\tau}a^{\dagger}_T a_T e^{-iH\tau}a_T|\psi\rangle}{|\langle\psi|a^{\dagger}_T a_T |\psi\rangle|^2},\label{g2}\\\nonumber
\end{eqnarray}

and similarly for the reflected field, where $|\psi\rangle$ is the steady-state wavevector. For stronger driving fields, we solve the full atomic density matrix equations, thus accounting for quantum jumps.

\section{LINEAR OPTICS}

\begin{figure}[b]
\centering
\includegraphics[width=0.64\textwidth]{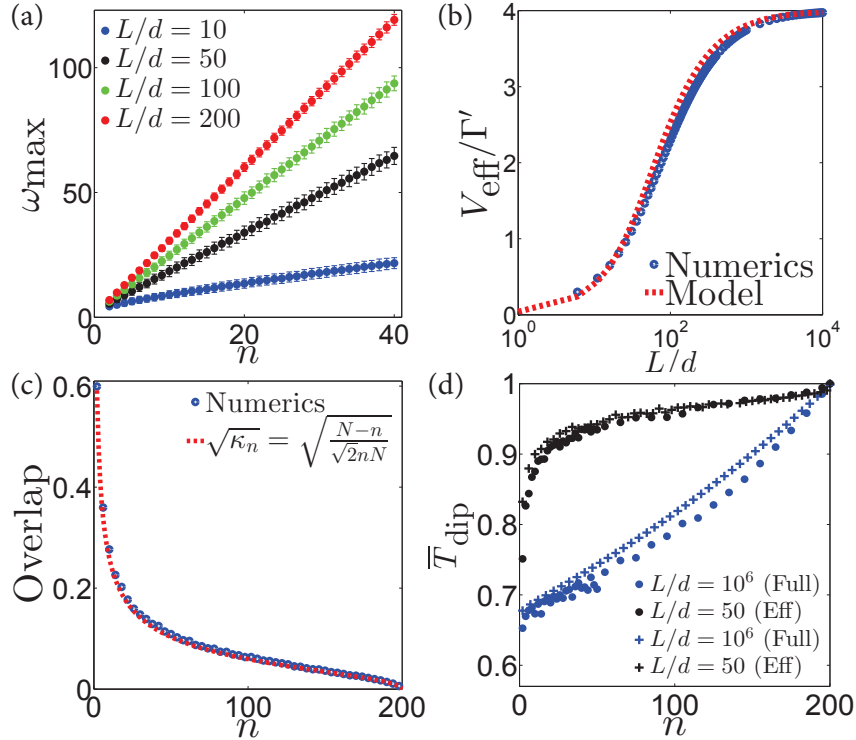}
\caption{(a) Mean (dots) and standard deviation (error bars) of $\omega_{\mbox{\scriptsize max}}$ vs. $n$ for different $L$, from 1000 samples of atomic positions per $n$ (other parameters as in Fig. 1). (b) $V_{\mbox{\scriptsize eff}}(L)$ calculated from Eq. (\ref{Veff}) (red dashed line) and from exact numerics (blue dots), taking $V/\Gamma^{\prime}=4$ and $N=200$, and $n\in[1,40]$. (c) Mean overlap (blue dots) of initial excitation and maximum-energy resonance vs. $n$, for the case $L/d=10^6$ and $k_Ld = \pi/2$; red dashed line shows a simple fit. (d) $\overline{T}_{\mbox{\scriptsize dip}}$ (see Fig. \ref{Fig1}d) vs. $n$ for the cases $L/d=10^6$ (blue) and $L/d=50$ (black) for the full and effective models, as described in the text.}
\label{Fig2}
\end{figure}

Fig. \ref{Fig1}c shows a representative single-shot (atoms fixed in position) linear transmission spectrum for one configuration of $n=10$ atoms randomly placed in a lattice of $N=200$ sites. We choose the lattice constant $d$ such that $k_ad=\pi/2$, which for unit filling factors (i.e. for $n/N=1$) minimizes back-reflection from the array, and thus most closely corresponds to a free-space atomic ensemble. However, different choices of $k_ad$ (excluding those very close to integer multiples of $\pi$) do not qualitatively change the results.

For the case $V=0$ (red curve in Fig. \ref{Fig1}c), the transmission spectrum is identical to that of a free-space ensemble, i.e. a broadened Lorentzian centered at the single-atom resonance frequency. The resonant ($\Delta=0$) transmittance is given by $T\sim e^{-D}$, where $D\approx2n\Gamma_{\g}/\Gamma^{\prime}$ is the optical depth. The blue curve shows the spectrum in the presence of the bandgap (BG) interaction, with strength $V/\Gamma^{\prime}=4$ and length $L/d=100$. Clearly, the system now has new resonance frequencies that may be resolved spectroscopically.

To understand the structure of the linear spectrum, note that in the limit $L\rightarrow\infty$, the BG interaction term in Eq. (\ref{Htot}) is described in the one-excitation manifold by an $n\times n$ matrix with entries $\pm V$, with the phase of a given element determined by the corresponding atomic positions. This matrix has $n-1$ degenerate resonances of energy equal to the bare single-atom energy, and one resonance of energy $nV$. For finite-range interactions, the $(n-1)$-fold degeneracy can be lifted to yield additional observable resonances, as shown in Fig. \ref{Fig1}c, while the maximum resonance energy, $\omega_{\mbox{\scriptsize max}}$, is correspondingly reduced. However, as we explain below, the properties of this maximum resonance can still be well quantified.

Current techniques for interfacing atoms with PCWs do not allow precise control of the atomic positions ${z_j}$, and moreover, due to finite trap lifetimes, single-shot spectroscopy is not possible. We therefore seek to understand the average optical properties, taking (unless stated otherwise) the mean of the appropriate quantities over 1000 random spatial configurations. Fig. \ref{Fig1}d shows the average linear transmission spectra for the same conditions as Fig. \ref{Fig1}c. While each realization has a unique value of $\omega_{\mbox{\scriptsize max}}$, as we explain in Appendices \ref{AppA1} \& \ref{AppA2}, for $L/d\gg1$ the average $\overline{\omega}_{\mbox{\scriptsize max}}$ scales linearly with the number of atoms in the system, as shown in Fig. \ref{Fig2}a, with a slope that is dependent on the ratio $L/Nd$. Indeed, fixing $Nd$ and defining an effective interaction strength $V_{\mbox{\scriptsize eff}}(L)$ such that $\overline{\omega}_{\mbox{\scriptsize max}}(L) = V + (n-1)V_{\mbox{\scriptsize eff}}(L)$, a simple estimate gives (see Appendix \ref{AppA1})

\begin{equation}\label{Veff}
V_{\mbox{\scriptsize eff}}(L) = \frac{2LV}{Nd}\left(1 - e^{-Nd/2L}\right).
\end{equation}

As shown in Fig. \ref{Fig2}b, this simple model agrees well with the numerically obtained solution. As we explain further in Appendix \ref{AppA1}, when $n\ll N$, $V_{\mbox{\scriptsize eff}}(L)$ may then be used to determine the frequency $\overline{\omega}_{\mbox{\scriptsize max}}$ for a PCW with known $V$, $L$, $N$, and $d$. It follows that the number of atoms may be inferred from the frequency of the highest-energy transmission minima. This may be a useful experimental characterization tool, given that the resonant optical depth ceases to be directly related to the total atom number in the presence of BG interactions.

Another important feature of the maximum resonance $\overline{\omega}_{\mbox{\scriptsize max}}$ to characterize is the associated transmittance $\overline{T}_{\mbox{\scriptsize dip}}$, as illustrated in Fig. 1d. Intuitively, this should be a function of the probability of creating the collective excitation associated with the maximum eigenvalue. In the limit $L\rightarrow\infty$, it may easily be shown that this excitation has the form $|\psi_{\mbox{\scriptsize max}}\rangle = (1/\sqrt{n})\sum_j(-1)^{\theta_j}|e_j\rangle$, with $\theta_j = z_j/d$. We can then quantify the probability of exciting this state by considering its overlap with the initial excitation $|\psi_{\mbox{\scriptsize in}}\rangle$ provided by the input field, where $|\psi_{\mbox{\scriptsize in}}\rangle = (1/\sqrt{n})\sum_j e^{ik_Lz_j}|e_j\rangle$. Fig. \ref{Fig2}c shows (as we explain fully in Appendix \ref{AppA3}) how, as a function of the number of atoms $n$, the mean overlap $\overline{\langle\phi_{\mbox{\scriptsize max}}|\psi_{\mbox{\scriptsize in}}\rangle}$ decays approximately as $1/\sqrt{n}$ for moderate filling factors, and tends to zero for very large filling factors. We find that  $\overline{\langle\phi_{\mbox{\scriptsize max}}|\psi_{\mbox{\scriptsize in}}\rangle}\sim\sqrt{\kappa_n}$, where $\kappa_n = (N-n)/(\sqrt{2}nN)$, gives a very good estimate of the overlap in all regimes.

\begin{figure}[h]
\centering
\includegraphics[width=0.8\textwidth]{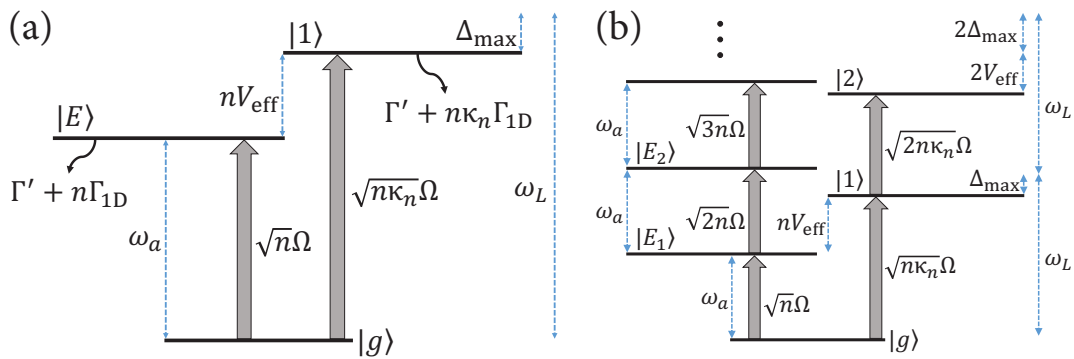}
\caption{(a) Level scheme and coupling and decay rates for an effective model in the linear regime. $|E\rangle$ represents a standard ensemble of atoms at the bare single-atom frequency, and $|1\rangle$ is the maximum single-photon resonance. (b) Extension to the many-photon regime: the ensemble now supports many excitations, and $|2\rangle$ is the maximum-energy resonance of the two-photon manifold. For clarity, the decay rates are omitted.}
\label{Fig3}
\end{figure}

As a result of the decreasing overlap, the transmittance $\overline{T}_{\mbox{\scriptsize dip}}$ increases as a function of $n$, as shown in Fig. \ref{Fig2}d (dots) for two different values of $L$. As explained fully in Appendix \ref{AppA3}, we can also construct an effective model that treats the system as having only two degrees of freedom - an `ensemble' level $|E\rangle$ with the bare single-atom resonance frequency, and the maximum resonance $|1\rangle$ at frequency $\overline{\omega}_{\mbox{\scriptsize max}}$, with coupling strengths and decay rates as shown in the level diagram of Fig. \ref{Fig3}a. The crosses in Fig. \ref{Fig2}d show the effective model predictions for $\overline{T}_{\mbox{\scriptsize dip}}$ for the two values of $L$, agreeing well with the full model in both cases. Moreover, for the case $L\rightarrow\infty$ the effective model may be solved exactly to give $\overline{T}_{\mbox{\scriptsize dip}}\sim \Gamma^{\prime^2}/(\Gamma^{\prime} + n\kappa_n\Gamma_{\g})^2$, which is analogous to the resonant transmission $T = \Gamma^{\prime^2}/(\Gamma^{\prime} + \Gamma_{\g})^2$ through a single atom coupled to the probe band \cite{chang2012cavity}.

\newpage
\section{NONLINEAR OPTICS}

We now turn to the nonlinear optical properties of the system, beginning with the second-order correlation function of the reflected field, $g^{(2)}_{R}$, when the system is driven at its maximum single-excitation energy, $\omega_{\mbox{\scriptsize max}}$. The reflected field arises purely due to scattering from the ensemble, whereas the transmitted field is an interference between the scattered and incident fields, see. Eq. (2). Thus, $g^{(2)}_{R}$ yields information about the ability of the system to scatter two photons simultaneously. The limiting case of a single two-level atom produces perfect anti-bunching, $g^{(2)}(0)=0$, indicating that it can only absorb and re-emit single excitations at a time.

\begin{figure}[t]
\centering
\includegraphics[width=0.8\textwidth]{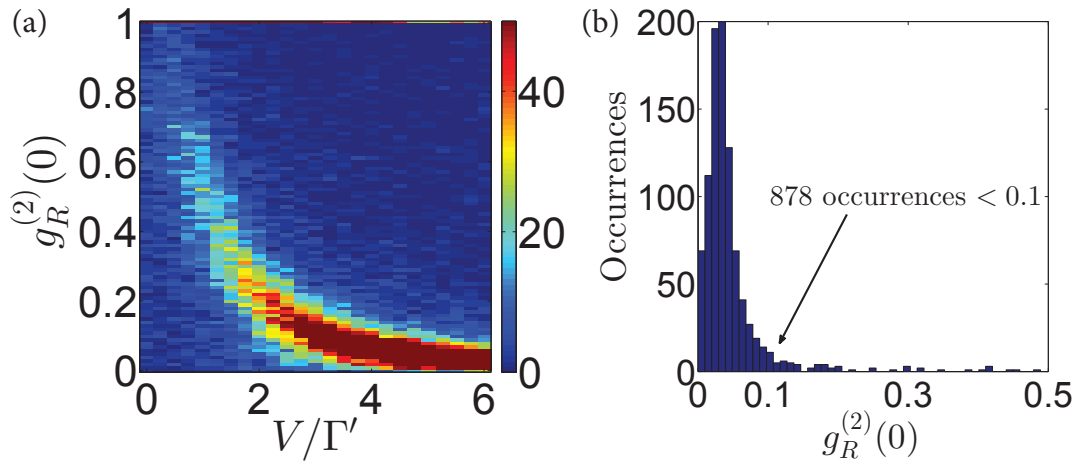}
\caption{(a) Number of occurrences (see color bar) of the y-axis value of $g^{(2)}_R(0)$ for given x-axis value of $V$, from 1000 samples of atomic positions per $V$. (b) Histogram of the case $V/\Gamma^{\prime}=6$. Other parameters: $n=20$, $N=200$, $L/d = 100$, $k_ad = \pi/2$, $\Omega/\Gamma^{\prime} = 0.01$, $\Gamma_{\g}/\Gamma^{\prime} = 0.3$.}
\label{Fig4}
\end{figure}

For a given value of $V$, we compute the correlation function $g^{(2)}_R(0)$ for each of 1000 random samples of atomic positions, plotting the occurrences of each value of $g^{(2)}_R(0)$ in Fig. \ref{Fig4}a. Evidently, the stronger the interaction, the more prevalent strong anti-bunching effects become. In particular, as Fig. \ref{Fig4}b shows, for an interaction strength $V/\Gamma^{\prime}=6$, the proportion of configurations with $g^{(2)}_R(0)<0.1$ approaches 90\%, suggesting that in the vicinity of the maximum resonance $\omega_{\mbox{\scriptsize max}}$, the ensemble typically behaves as an effective two-level system.

The origin of the anti-bunching in the $L\rightarrow\infty$ limit may easily be understood: by diagonalization of the bandgap interaction (proportional to $V$ in Eq. (1)), we find the maximum single-photon eigenenergy to be $\omega_{\mbox{\scriptsize max}}^{(1)}= nV$, and the maximum two-photon eigenenergy to be $\omega_{\mbox{\scriptsize max}}^{(2)}=2(n-1)V$. Defining the anharmonicity $A = \omega_{\mbox{\scriptsize max}}^{(2)}-2\omega_{\mbox{\scriptsize max}}^{(1)}$, we find $A(L=\infty) = -2V$. Thus, it is clear that for large V the system approaches ideal two-level behavior. For finite $L$, provided that the linear spectrum is well characterized by $V_{\mbox{\scriptsize eff}}(L)$ (Eq. \ref{Veff}), the anharmonicity $A(L) \sim -2V_{\mbox{\scriptsize eff}}(L)$ - see Appendix \ref{AppB2}.

Motivated by the above observation, an interesting question is whether a global Rabi pulse applied to all of the atoms can produce only a single collective excitation in the ensemble (as well as the associated fidelity as a function of the various system and laser parameters). To answer this, we study the dynamics under Eq. (\ref{Htot}) with a driving field $\Omega$ that is sufficiently large to produce Rabi oscillations, solving the full master equation for the atomic density matrix when the system is initialized in its ground state $|g\rangle^{\otimes n}$.

\begin{figure}
\centering
\includegraphics[width=0.65\textwidth]{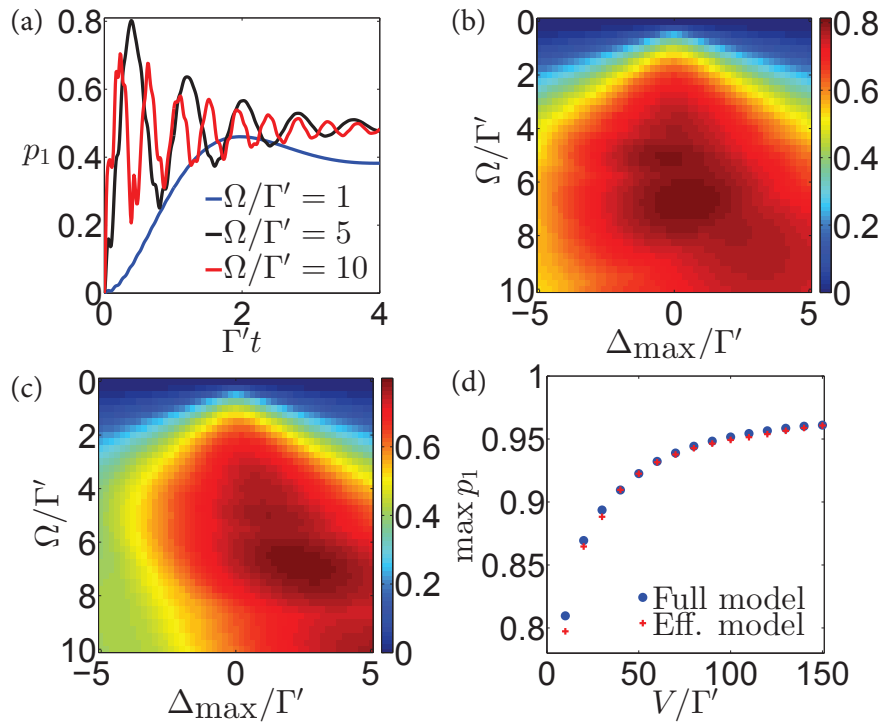}
\caption{(a) Time-dependent probability $p_1$ for an atomic ensemble to contain exactly one excitation, for various Rabi frequencies $\Omega$. For certain values of $\Omega$, the probability $p_1$ can become very high, indicating that the ensemble effectively behaves as a two-level system. The other parameters used for this plot are $n=6$, $V/\Gamma^{\prime}=10$, $L/d=10^6$, $N=50$, $\Delta_{\mbox{\scriptsize max}}/\Gamma^{\prime} = 0$. The atomic positions were chosen to give overlap $\langle\phi_{\mbox{\scriptsize max}}|\psi_{\mbox{\scriptsize in}}\rangle \approx\sqrt{\kappa_6}$. (b) The maximum single-excitation population $p_1$ achieved in time, as a function of detuning $\Delta_{\mbox{\scriptsize max}}$ and laser strength $\Omega$ for the same configuration as (a). (c) The maximum single-excitation population $p_1$ achieved in time, as predicted from a simplified effective model for the same conditions as (b). (d) $\max p_1$ as a function of $V$ as predicted by the full and effective models (the full model for the same $n=6$ configuration as in (a)). In all cases, $k_ad = \pi/2$ and $\Gamma_{\g}/\Gamma^{\prime} = 0.3$.}
\label{Fig5}
\end{figure}

Fig. \ref{Fig5}a shows a representative sample of the time evolution of the single-excitation manifold population, $p_1$, for a system of 6 atoms driven at its maximum resonance frequency and for different driving strengths, with $V/\Gamma^{\prime}=10$. Clearly, depending on the driving strength, $p_1$ can be very large. Intuitively, to maximize $p_1$, $\Omega$ must be sufficiently large to overcome dissipation, yet sufficiently small to minimize subsequent population of the doubly-excited manifold: we observe this trade-off in Fig. \ref{Fig5}a, where the choice $\Omega/\Gamma^{\prime}=5$ gives a larger maximum value of $p_1$ than the cases $\Omega/\Gamma^{\prime}=1$ and $\Omega/\Gamma^{\prime}=10$.

Fig. \ref{Fig5}b shows the maximum value of $p_1$ during the evolution time, as a function of both $\Omega$ and the detuning $\Delta_{\mbox{\scriptsize max}} = \omega_L - \omega_{\mbox{\scriptsize max}}$ from the maximum resonance frequency. Since the anharmonicity is negative, to minimize the probability of two-photon absorption the optimal choice of $\Delta_{\mbox{\scriptsize max}}$ is positive. The optimal single-excitation probability $p_1^{\mbox{\scriptsize opt}} \approx 0.81$ occurs for the parameters $\Omega^{\mbox{\scriptsize opt}}/\Gamma^{\prime} = 6.75$, $\Delta_{\mbox{\scriptsize max}}^{\mbox{\scriptsize opt}}/\Gamma^{\prime} = 0.4$.

While Figs. \ref{Fig5}a,b are calculated with full density matrix dynamics, we now introduce a simplified effective model, building on the linear model discussed above. This enables both a simple understanding of the dependence of the maximum single excitation probability $p_1$ versus system parameters, and predictions when the atom number becomes too large for exact density matrix solutions to be feasible. As depicted in Fig. \ref{Fig3}b, the model treats the system as having two separate `branches', corresponding to the ensemble of independent atoms (with $m^{th}$ excited state $|E_m\rangle$), and the spin-wave excitation of the maximum-energy resonances in the one- and two-photon manifolds, denoted by states $|1\rangle$ and $|2\rangle$.

Fig. \ref{Fig5}c shows the effective model prediction for $p_1$ as a function of the external field properties, under the same conditions as for Fig. \ref{Fig5}b. While the optimal laser parameters ($\Omega^{\mbox{\scriptsize opt}}/\Gamma^{\prime} = 6.75$ and $\Delta_{\mbox{\scriptsize max}}^{\mbox{\scriptsize opt}}/\Gamma^{\prime} = 2.4$) do not match those of the full model, the optimal single-excitation probability is accurate to within $\sim 1\%$ ($p_1^{\mbox{\scriptsize opt}} \approx 0.80$). We may then consider the fidelity of introducing only a single excitation for given physical resources $n$ and $V$. Fig. \ref{Fig5}d shows the maximum value of $p_1$ obtained for the case $n=6$ as a function of $V$, with both models predicting that $p_1$ approaches unity for large $V$.

Using the effective model, we may analytically obtain an estimate of the error in introducing a single excitation, as a result of population in the second-excited manifold. As we explain in Appendix \ref{AppB4}, in the limit $\Omega\ll\sqrt{n}V_{\mbox{\scriptsize eff}}$, where the ensemble may be neglected, the dominant error comes from the finite anharmonicity of the maximum resonance, which yields a small population in state $|2\rangle$ of $p_{|2\rangle}\sim \Omega_n^2/4V^2$, where $\Omega_n = \sqrt{n\kappa_n}\Omega$. The total error is then well-approximated by a sum of the error in creating perfect population inversion for an ideal two-level system (under the conditions $\Omega_n$, $\Delta_{\mbox{\scriptsize max}}$, $\Gamma_{\g}$ and $\Gamma^{\prime}$), plus the contribution from $p_{|2\rangle}$. While there is no simple closed-form expression, the total error can nonetheless be easily optimized numerically (see Appendix \ref{AppB4}).

\section{SUMMARY AND OUTLOOK}

In summary, we have shown that an ensemble of two-level atoms with long-range interactions has rich optical properties in both the single- and many-photon regimes. We have considered the simplest conceptual model, primarily in order to understand and inform experiments that are within reach. Moving forward, it may be possible to exploit strong nonlinearities to produce a blockade effect, where within some region of the system only a single excitation is supported. If the full system is large enough to support several such regions, one would expect the scattered light to exhibit rich spatiotemporal correlations \cite{DarrickNatureReview}.

\section{ACKNOWLEDGEMENTS}

We thank J. S. Douglas for stimulating discussions. EM and LCK acknowledge support from the National Research Foundation \& Ministry of Education, Singapore. DEC acknowledges support from the ERC Starting Grant FOQAL, MINECO Plan Nacional Grant CANS, and MINECO Severo Ochoa Grant SEV-2015-0522, and Fundacio Privada Cellex.

\bibliography{references}
\bibliographystyle{unsrt}

\newpage
\appendix

\section{LINEAR OPTICS}
\setcounter{figure}{0}
\renewcommand{\thefigure}{A\arabic{figure}}
\renewcommand{\theHfigure}{A\arabic{figure}}
\setcounter{equation}{0}
\renewcommand{\theequation}{A\arabic{equation}}%

\subsection{Effective interaction strength}\label{AppA1}

Figure \ref{Fig2}a in the main text shows that for finite values of the characteristic interaction length $L$, the average energy of the highest single-excitation resonance $\overline{\omega}_{\mbox{\scriptsize max}}$ scales linearly with the number of atoms $n$ in the system. This motivates the definition of an effective interaction strength $V_{\mbox{\scriptsize eff}}(L)$, through the equation $\overline{\omega}_{\mbox{\scriptsize max}}(L) = V + (n-1)V_{\mbox{\scriptsize eff}}(L)$. The first factor of $V$ is the energy of a single atom, and its average interaction energy with each subsequent atom is $V_{\mbox{\scriptsize eff}}(L)$. Here we give a simple model for estimating $V_{\mbox{\scriptsize eff}}$ for a photonic crystal of given $V$, $L$, and a lattice of $N$ sites separated by distance $d$.

The band-gap interaction Hamiltonian is

\begin{equation}
H_{BG} = V\sum_{j,k} e^{-|z_j - z_k|/L}\sigma_{eg}^j\sigma_{ge}^k,
\end{equation}

where, for simplicity, we neglect the phase factor resulting from the product of Bloch wavefunctions. When the scaling is linear, it is sufficient to consider the average interaction energy of two atoms, in which case the eigenstates of $H_{BG}$ in the single-photon manifold are simply the symmetric and anti-symmetric states $|\pm\rangle = (1/\sqrt{2})(|g\rangle \pm |e\rangle)$, and we may easily determine the corresponding energies. Fixing the first atom to be in the middle of the lattice (i.e. in its average position), the energy of the symmetric state $|+\rangle$ (the state of higher energy), averaged over all possible separations, is

\begin{equation}\label{VeffApp}
V_{\mbox{\scriptsize eff}} = \frac{2V}{N}\sum_{j=1}^{N/2} e^{-jd/L} = \frac{2V}{N}\frac{1 - e^{-Nd/2L}}{e^{d/L} - 1} \sim \frac{2LV}{Nd}\left(1 - e^{-Nd/2L}\right).
\end{equation}

where we assume $L\gg d$. As shown in Fig. \ref{Fig2}b of the main text, this simple model agrees well with the results from exact numerics, however we now consider the limits of its applicability.

\begin{figure}[H]
\centering
\subfigure[]{
\label{omegamax_Veff_LargerL}
\includegraphics[width=0.45\textwidth]{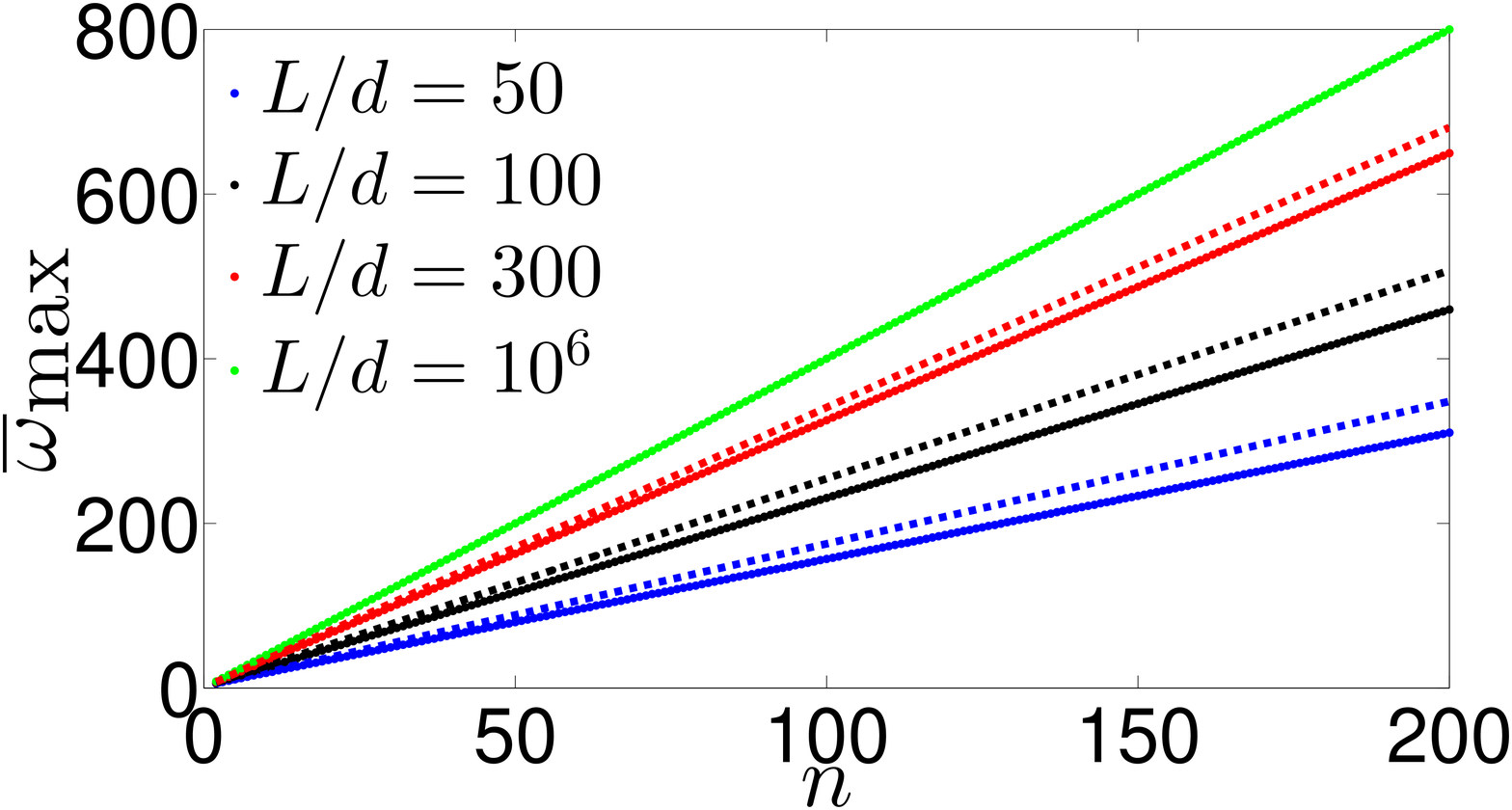}}
\subfigure[]{
\label{diff_slope_Veff}
\includegraphics[width=0.45\textwidth]{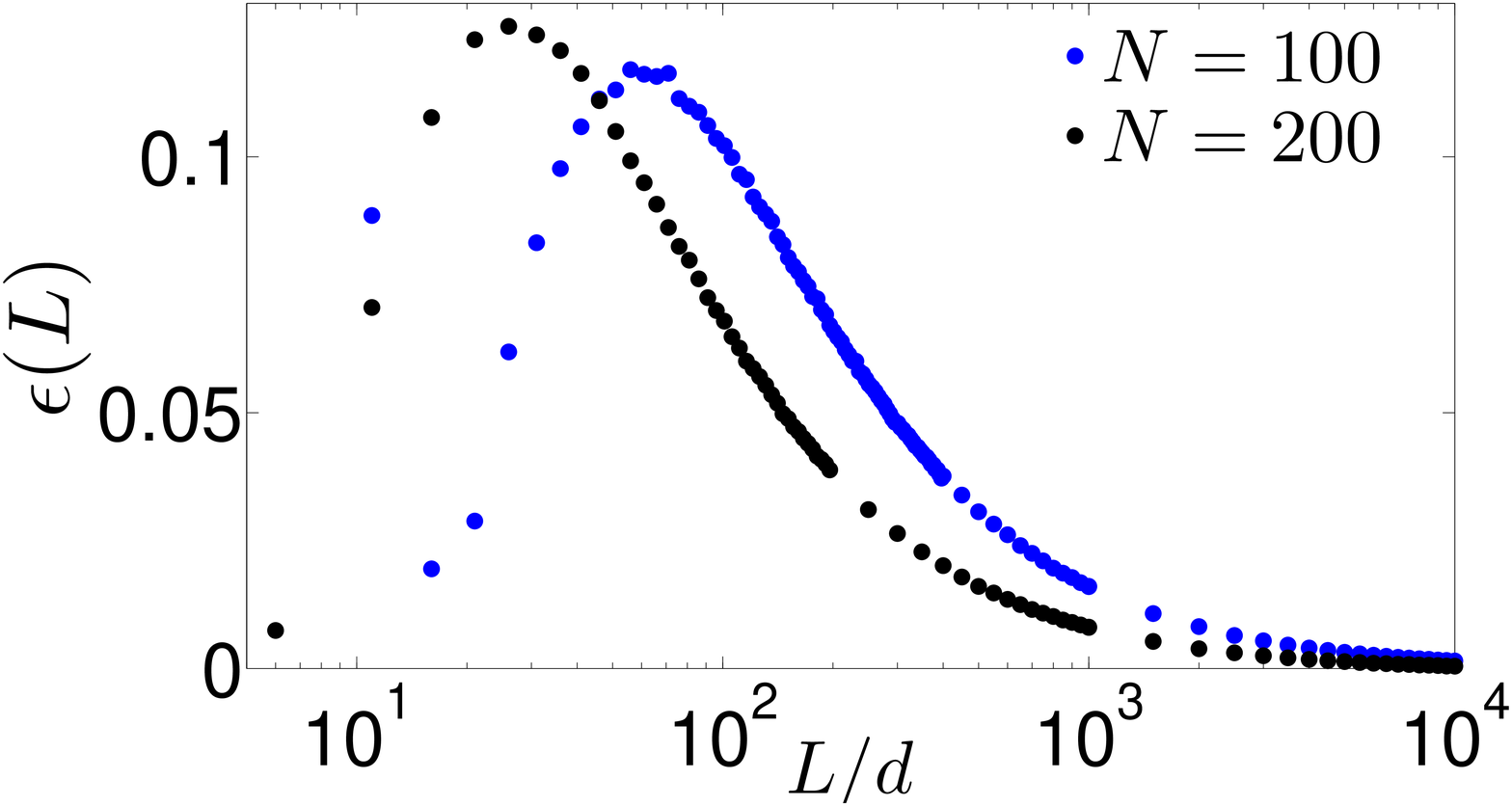}}
\caption{(a) Average frequency $\overline{\omega}_{\mbox{\scriptsize max}}$ vs. number of atoms $n$, showing that the approximation using $V_{\mbox{\scriptsize eff}}$ becomes less accurate as $n\rightarrow N$ for $L\neq\infty$. Here, $V/\Gamma^{\prime}=4$, $\Gamma_{\g}/\Gamma^{\prime}=0.3$, and $k_ad = \pi/2$. 1000 samples were taken per $n$, for a system of total length $N=200$ sites. (b) Error $\epsilon(L)$ in the slope $V_{\mbox{\scriptsize eff}}$ relative to the true slope of $\overline{\omega}_{\mbox{\scriptsize max}}$ vs. $n$.}
\label{Veff_slopes}
\end{figure}

The derivation of Eq. \ref{VeffApp} makes use of the fact that we know the maximum-energy eigenstate to be $|+\rangle$, in which the excitation is equally distributed between the two atoms. The assumption that we may use the same expression for the energy in the case of $n$ atoms therefore assumes in turn that the $n$-atom eigenstate is of the form $|\psi_n\rangle = (1/\sqrt{n})\sum_j|e_j\rangle$, which is strictly true only for the case $L/d=\infty$. The approximation using $V_{\mbox{\scriptsize eff}}$ over-estimates the slope of the linear relationship between $\overline{\omega}_{\mbox{\scriptsize max}}$ and $n$, as is shown in Fig. A\ref{omegamax_Veff_LargerL} for several different cases. The accuracy of the approximation therefore diminishes as $n$ increases.

To gain further insight into the validity of the model, we consider the relative error $\epsilon(L)$ in the slope $V_{\mbox{\scriptsize eff}}$, defined as

\begin{equation}
\epsilon(L) = \frac{|V_{\mbox{\scriptsize eff}}(L) - S(L)|}{S(L)},
\end{equation}

where S(L) is the true slope of the maximum resonance frequency vs. $n$ line, as computed using the full model. Fig. A\ref{diff_slope_Veff} shows this error as a function of $L$, for the cases $N=100$ and $N=200$. As one would expect on the basis of the above discussion, $\epsilon(L)$ tends to zero as $L\rightarrow\infty$. In both cases, the maximum error occurs for $L/d \approx N/2$.

\subsection{Breakdown of linear scaling of $\overline{\omega}_{\mbox{\scriptsize max}}$ with $n$}\label{AppA2}

For interaction lengths $L/d\gg1$ the scaling of $\overline{\omega}_{\mbox{\scriptsize max}}$ with $n$ is linear, and moreover in the limit $n\ll N$, $\overline{\omega}_{\mbox{\scriptsize max}}$ is well characterized by $V_{\mbox{\scriptsize eff}}$. In contrast, for very short range interactions ($L/d\sim1$) the coupling between distant atoms can become negligible, resulting in a total energy that scales sub-linearly in the number of atoms $n$. As a simple example, consider the case of $n=4$ atoms in a lattice of $N=200$ sites, with an interaction length $L/d \sim \mathcal{O}(1)$. One possible configuration of atomic positions in this system is $z_j = (1,2,199,200)$, where the atoms form two disconnected `blocks'. In this instance, the maximum resonance energy is proportional to that of two interacting atoms, rather than four.

\begin{figure}[H]
\centering
\includegraphics[width=0.45\textwidth]{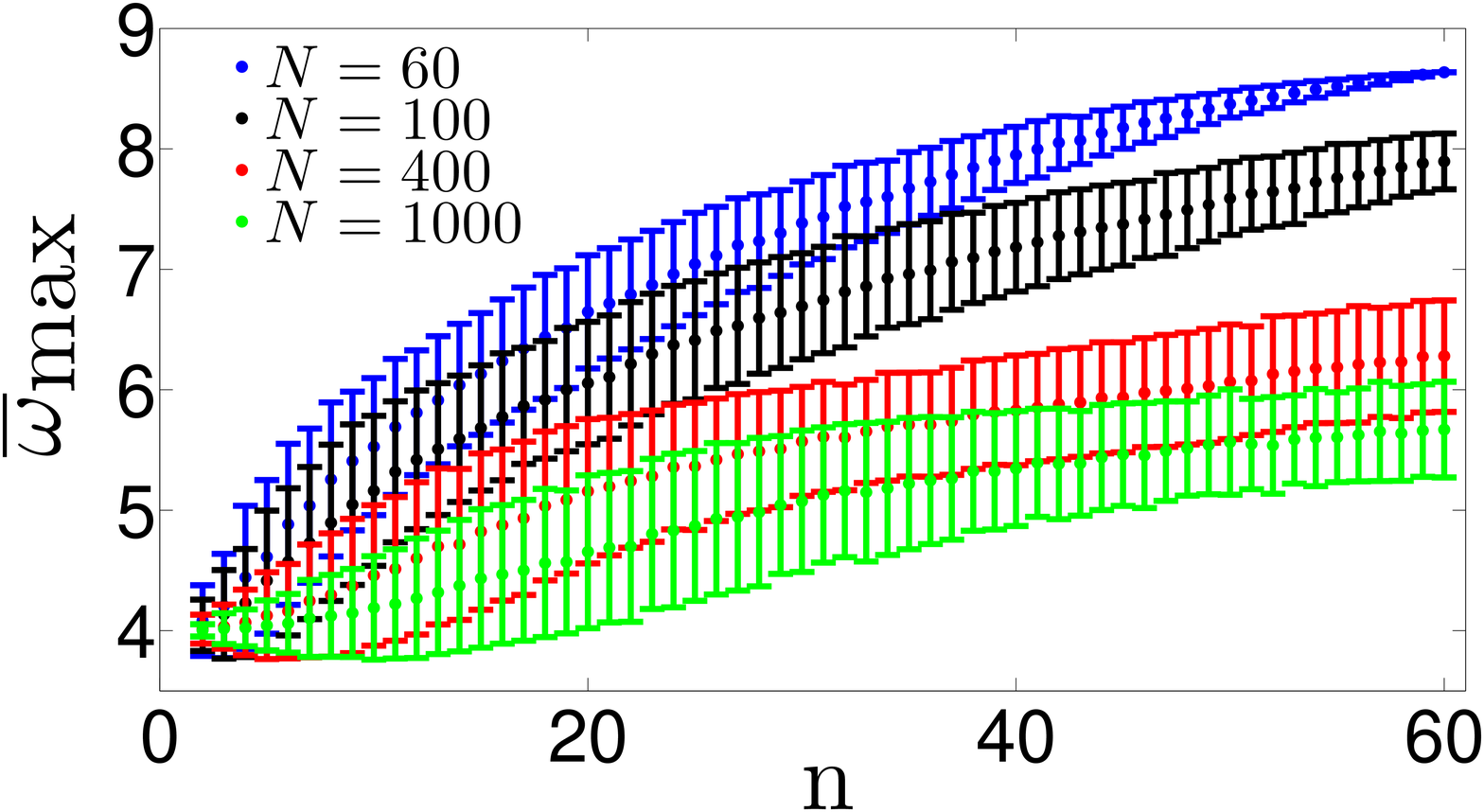}
\caption{$\overline{\omega}_{\mbox{\scriptsize max}}$ vs. $n$ for the case $L/d=1$ and systems of different lengths, with all other parameters as in Fig. A\ref{omegamax_Veff_LargerL}.}
\label{BGres_functionatoms_L1}
\end{figure}

Figure \ref{BGres_functionatoms_L1} shows how the mean maximum resonance energy varies with $n$ for the case $L/d=1$, for different total system sizes. The energy is largest (smallest) in the case $N=60$ ($N=1000$) since the average separation between atoms is shortest (longest). In all cases, the rate of increase in energy diminishes as the system filling increases, in line with the fact that atoms that are increasingly distant couple increasingly weakly.

\subsection{Overlap of highest resonance with initial excitation}\label{AppA3}

The TM-mode probe field creates an initial excitation in the system with a phase profile determined by the positions of the atoms. We write this initial state as $|\psi_{\mbox{\scriptsize in}}\rangle = \frac{1}{\sqrt{n}}\sum_j e^{ik_Lz_j}|e_j\rangle$, where $|e_j\rangle$ denotes an excitation at the $j^{th}$ atom, with all others in the ground state. For the choice $k_Ld = \pi/2$ this gives $|\psi_{\mbox{\scriptsize in}}\rangle = \frac{1}{\sqrt{n}}\sum_j (i)^{\theta_j}|e_j\rangle$, with $\theta_j =z_j/d$. Meanwhile, in the limit $L\rightarrow\infty$, it may easily be shown that the maximum-energy resonance of the BG interaction has the form $|\phi_{\mbox{\scriptsize max}}\rangle = \frac{1}{\sqrt{n}}\sum_j (-1)^{\theta_j}|e_j\rangle$, i.e. where each dipole oscillates $\pi$ out of phase with its nearest neighbour. The overlap $\kappa$ of the two states is then

\begin{equation}
\kappa \equiv \langle\psi_{\mbox{\scriptsize in}}|\phi_{\mbox{\scriptsize max}}\rangle = \left(\frac{1}{\sqrt{n}}\sum_j\langle e_j|(i)^{\theta_j}\right)\left(\frac{1}{\sqrt{n}}\sum_k(-1)^{\theta_k}|e_k\rangle\right) = \frac{1}{n}\sum_j(-i)^{\theta_j + \theta_k}.
\end{equation}

Since the atomic positions are non-deterministic for any given realization, the latter equality shows that the overlap is proportional to a sum of random phases along the real and imaginary directions. By analogy to random walks, statistically - i.e in taking many samples and averaging - the magnitude of this sum scales as $\sim\sqrt{n/\sqrt{2}}$, from which it follows that the mean overlap $\overline{\kappa}$ scales as $1/\sqrt{\sqrt{2}n}$. Only in the special case $k_Ld\approx\pi$ does the summation of phases become coherent, where the $n$ dependence cancels and $\overline{\kappa}\rightarrow 1$.

The above reasoning breaks down as the number of atoms approaches the total number of lattice sites, since then the sum contains an equal number of terms of opposite phases and thus tends to zero. For example, for $n=N=200$, there are 50 terms of each of the phases $1,i,-1,-i$, which when summed give a vanishing overlap. As shown in Figure \ref{Fig2}c in the main text, we find that the simple formula $\sqrt{\kappa_n} = \sqrt{(N-n)/\sqrt{2}nN}$ agrees very well with the numerically obtained scaling for all $n$.

\begin{figure}[H]
\centering
\includegraphics[width=0.45\textwidth]{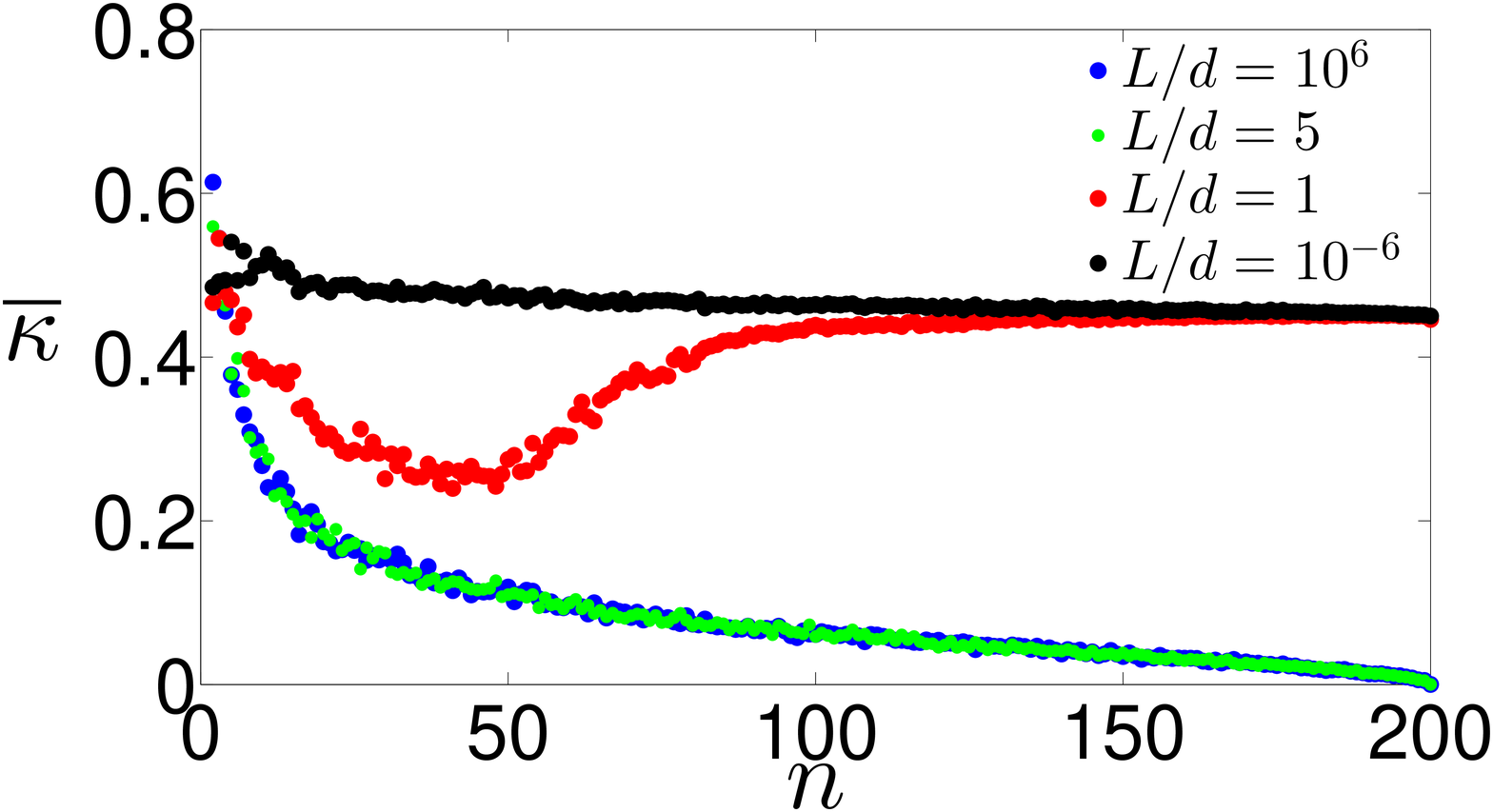}
\caption{Mean overlap $\overline{\kappa} = \overline{\langle\psi_{\mbox{\scriptsize in}}|\phi_{\mbox{\scriptsize max}}\rangle}$ for different interaction lengths, and a lattice of $N=200$ sites with $V/\Gamma^{\prime}=4$, $\Gamma_{1D}/\Gamma^{\prime}=0.3$.}
\label{overlap_atoms_differentL_N200}
\end{figure}

The above reasoning also assumes that $L\rightarrow\infty$, however, as  Fig. \ref{overlap_atoms_differentL_N200} shows for a lattice of $N=200$ sites, the same scaling behavior for $\overline{\kappa}$ holds for interactions with $L/d\gtrsim1$. Below this limit, the BG interaction is no longer dominant and thus no longer determines the maximum-energy resonance.

\subsection{Effective model for transmittance at maximum resonance}\label{LinearToyModel}\label{AppA4}

Here we bring together the key features of the maximum-energy resonance that we have found in the linear regime, constructing an effective model that aims to reproduce the results of the full model of Eq. \ref{Htot} in the main text. Specifically, we consider the transmittance $\overline{T}_{\mbox{\scriptsize dip}}$ as illustrated in Fig. \ref{Fig1}d of the main text. We define the detuning of the probe field from the maximum-energy resonance to be $\Delta_{\mbox{\scriptsize max}} = \omega_L - \omega_{\mbox{\scriptsize max}}$. The ingredients for the model, assuming $L/d$ to be sufficiently large, are:

\begin{itemize}
\item An ensemble of $n$ independent atoms at the frequency $-(\Delta_{\mbox{\scriptsize max}} + nV_{\mbox{\scriptsize eff}}(L))$, denoted by a single level $|E\rangle$, whose coupling to the input field is enhanced by a factor of $\sqrt{n}$, and whose decay into the guided modes is enhanced by a factor of $n$. This approximately captures the response of a normal atomic ensemble driven far from resonance.
\item A single degree of freedom $|1\rangle$ at frequency $\Delta_{\mbox{\scriptsize max}}=0$ representing the maximum-energy resonance. In order to determine the coupling of the input field to this state, we note that the matrix element for coupling of the field to the initial atomic excitation $|\psi_{\mbox{\scriptsize in}}\rangle$ is $\sim\sqrt{n}\Omega$, while the subsequent overlap with the maximum-energy resonance is given by $\sim \sqrt{\kappa_n}$, as described in the previous section. Overall, therefore, the matrix element coupling the input field to the maximum resonance is $\sqrt{n\kappa_{n}}\Omega$.
\item For $L/d\rightarrow\infty$ the frequency of the maximum resonance is independent of the atomic positions, whereas for finite $L/d$ each configuration has a different value of $\omega_{\mbox{\scriptsize max}}$. As a result, in the averaged spectrum there is less attenuation at a given frequency in the latter case than the former, and hence the transmittance is higher. For the case $N=200$, $L/d=50$ presented in the main text, numerical analysis shows the spread of $\omega_{\mbox{\scriptsize max}}$ to be well approximated by a Gaussian distribution, so that for given values of $L$ and $n$ we can obtain the standard deviation $\sigma(n,L)$, which we then input this to the model via a random number $\eta$ with the appropriate statistics. Finally, we average over many samples of different $\eta$.
\end{itemize}

Figure \ref{Fig3}a in the main text schematically illustrates the level diagram for the model, with the various frequencies, couplings, and decay rates. The Hamiltonian is

\begin{eqnarray}
H_{\mbox{\scriptsize eff}} &=& -\left(\Delta_{\mbox{\scriptsize max}} + nV_{\mbox{\scriptsize eff}}(L) + \frac{i(\Gamma^{\prime} + n\Gamma_{\g})}{2}\right)\sigma_{EE} - \left(\Delta_{\mbox{\scriptsize max}} - \eta + \frac{i(\Gamma^{\prime} + n\kappa_n\Gamma_{\g})}{2}\right)\sigma_{11}\\ \nonumber
&& -\sqrt{n}\Omega\left(\sigma_{Eg} + h.c.\right) - \sqrt{n\kappa_{n}}\Omega\left(\sigma_{1g} + h.c.\right).
\end{eqnarray}

Here, as described above, $\eta$ is a random number of zero mean and standard deviation $\sigma(n,L)$. We emphasize that we do not rigorously derive this effective Hamiltonian from the full Hamiltonian of Eq. \ref{Htot} in the main text, but rather it represents a minimal model meant to capture the most important characteristics of the system. The output operator for the transmitted field is

\begin{equation}
a_T = \Omega  + \frac{i\Gamma_{\g}}{2}\left(\sqrt{n}\sigma_{gE} + \sqrt{n\kappa_{n}}\sigma_{g1}\right) + F,
\end{equation}

where the noise operator $F$ describes quantum jump processes, which may be neglected for weak driving ($\Omega/\Gamma^{\prime}\ll 1$). Using this model we obtain the results of Fig. \ref{Fig2}d (crosses) in the main text, taking 1000 samples of different $\eta$ in the case $L/d=50$.

In the case $L\rightarrow\infty$, we can also use the effective model to analytically obtain a simple formula for the transmittance $\overline{T}_{\mbox{\scriptsize dip}}$. Assuming $\sqrt{n}V_{\mbox{\scriptsize eff}}\gg\Omega,n\Gamma_{\g}$, the ensemble $|E\rangle$ may be neglected and the problem is reduced to that of a two-level system comprising the levels $|g\rangle, |1\rangle$. Solving the corresponding optical Bloch equations in the steady state (SS), we find that the transmittance is given by

\begin{eqnarray}
\overline{T}_{\mbox{\scriptsize dip}} &=& \frac{\langle a_T^{\dagger}a_T\rangle_{SS}}{\Omega^2} = \frac{\Gamma^{\prime^2}}{\left(\Gamma^{\prime} + n\kappa_n\Gamma_{\g}\right)^2}.
\end{eqnarray}

\subsection{Spectra for random number of atoms}\label{AppA5}

\begin{figure}[H]
\centering
\subfigure[]{
\label{poissonaveraging_4atoms}
\includegraphics[width=0.45\textwidth]{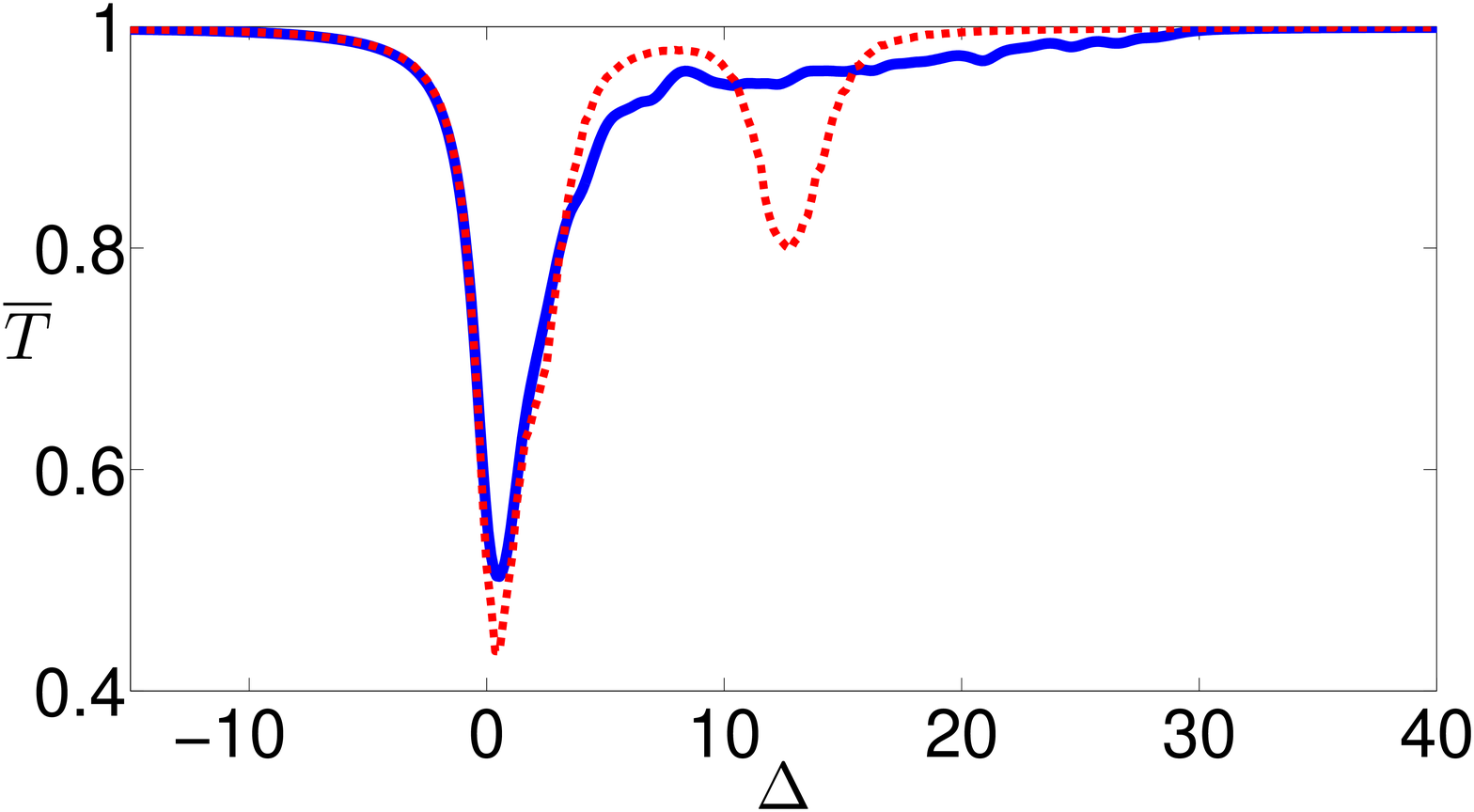}}
\subfigure[]{
\label{poissonaveraging_15atoms}
\includegraphics[width=0.45\textwidth]{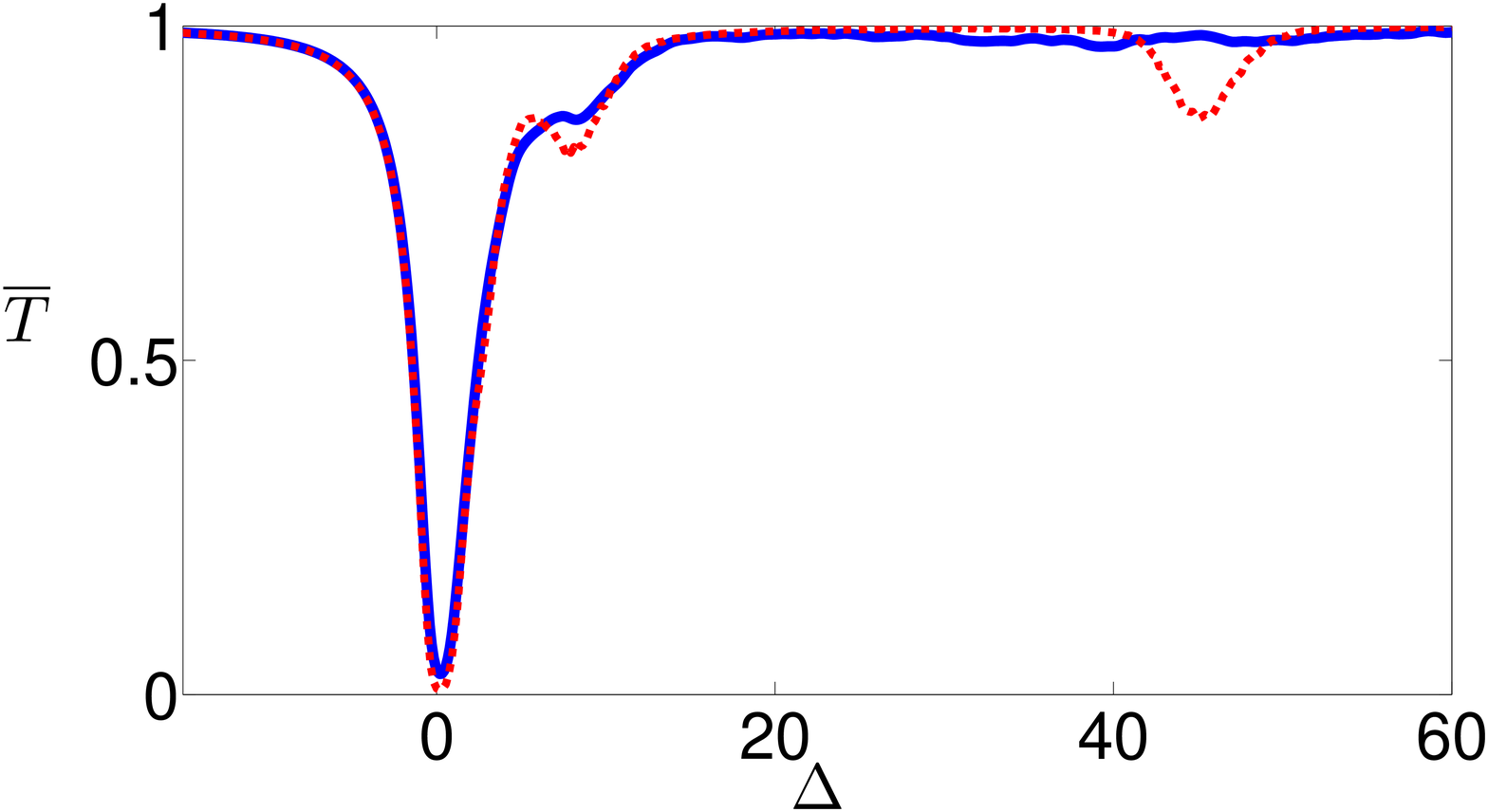}}
\caption{Transmittance spectra averaged over both atom number and positions (solid blue lines), and averaged only over positions for fixed atom number (red dashed line). Parameters used: $V/\Gamma^{\prime}=4$, $N=200$, $L/d=N$, 200 samples, for average atom numbers of (a) $\langle n \rangle = 4$ and (b) $\langle n \rangle = 15$.}
\label{poissonaveraging}
\end{figure}

In addition to averaging over atomic configurations for a fixed number of atoms, we can also take into account the effects of fluctuations in the number of atoms in the system. For a given average number of atoms $\langle n\rangle$ we use a Poisson distribution to give a random number $m$ of atoms, generate a random configuration of positions of these $m$ atoms, then obtain the transmission spectrum; we then repeat many times and take the average. As figure \ref{poissonaveraging} shows, the implication for the maximum resonance is that it can now occur over an even larger range of energies, so that in the averaged transmission spectra its signature is flattened out.

\section{NONLINEAR OPTICS}
\setcounter{figure}{0}
\renewcommand{\thefigure}{B\arabic{figure}}%
\renewcommand{\theHfigure}{B\arabic{figure}}
\setcounter{equation}{0}
\renewcommand{\theequation}{B\arabic{equation}}%

\subsection{Anti-bunching at other resonances}\label{AppB1}

\begin{figure}[H]
\centering
\includegraphics[width=0.5\textwidth]{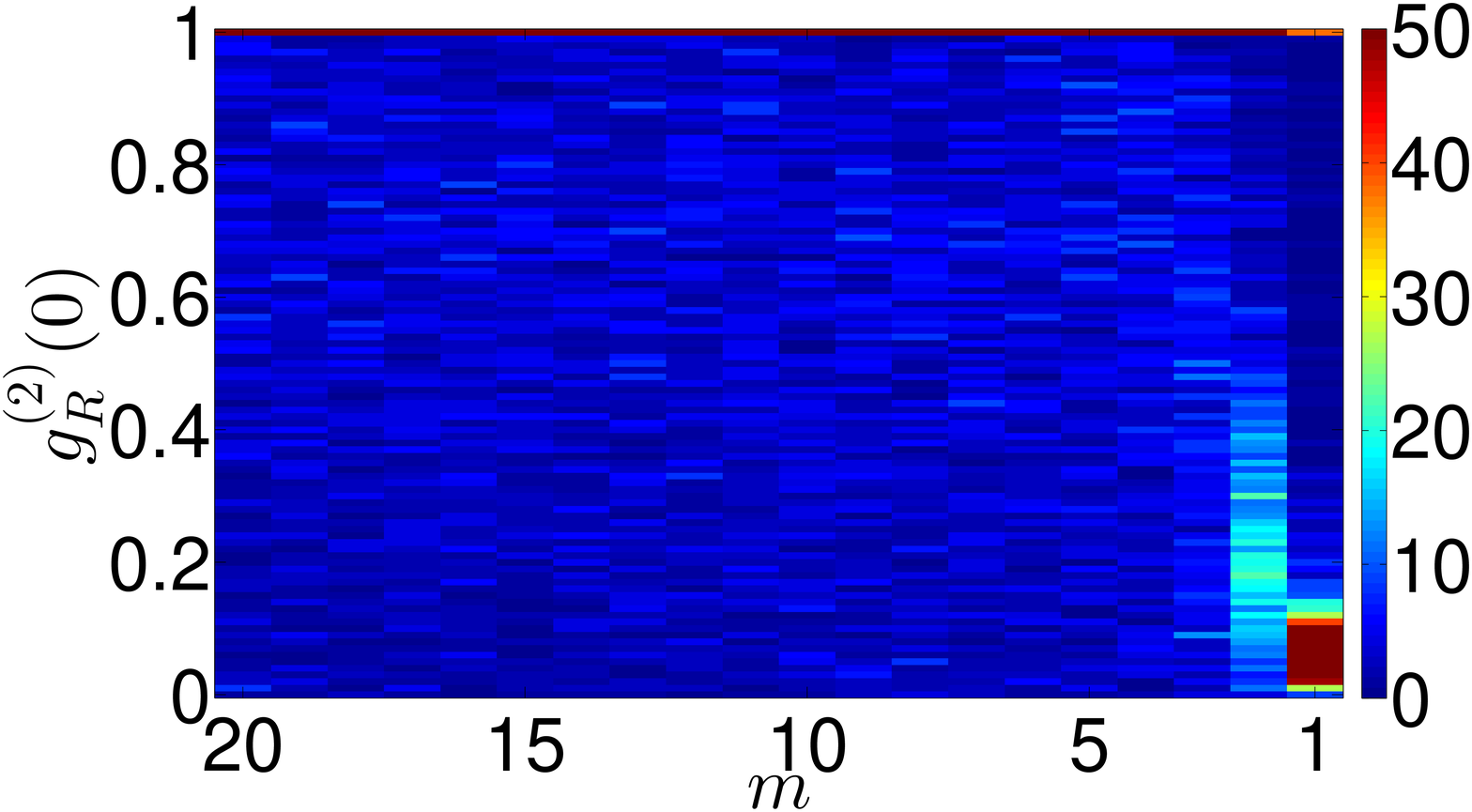}
\caption{Colormap histogram (from 1000 samples of atomic positions) of $g^{(2)}_R(0)$ for driving at the different resonance frequencies of the single-photon manifold ($m$ labels these resonances). Other parameters: $V/\Gamma^{\prime}=4$, number of atoms $n$ (sites $N$) $= 20$ $(200)$, $L/d=100$.}
\label{colormapV4hierarchy}
\end{figure}

Figure \ref{Fig4} in the main text shows that there is a very high probability of observing strong photon anti-bunching in the reflected-field correlation function $g^{(2)}_R(0)$ for $V>\Gamma^{\prime}$ when the system is driven at its highest single-photon resonance frequency. It is natural to then consider whether the same is true when the driving is instead at the frequencies of other resonances in the single-photon manifold. To investigate this, we diagonalize the interaction Hamiltonian $H_{BG}$ in the single-excitation manifold to obtain a set of (in general non-degenerate) resonances, which we index by $m=1,...,n$ from highest to lowest energy (i.e. $m=1$ is the maximum-energy eigenstate). We then weakly drive the system at each frequency $\omega_m$ and compute the correlation function $g^{(2)}_R(0)$. Again, we consider 1000 samples of atomic positions, and in Fig. \ref{colormapV4hierarchy} plot a histogram of the occurrences of given values of $g^{(2)}_R(0)$ (along the $y$-axis) for each $m$. Evidently, only the $m=1$ resonance gives rise to strong anti-bunching in a large proportion of cases.

\subsection{Anharmonicity for finite $L$}\label{AppB2}

In the limit $L\rightarrow\infty$ we argued in the main text that the anharmonicity - defined as $A = \omega_{\mbox{\scriptsize max}}^{(2)}-2\omega_{\mbox{\scriptsize max}}^{(1)}$, where $\omega_{\mbox{\scriptsize max}}^{(2)}$ ($\omega_{\mbox{\scriptsize max}}^{(1)}$) is the highest eigenvalue of the second (first) excited manifold -  is given by $A(\infty) = -2V$. For finite-range interactions and $n\ll N$, as discussed in Appendices \ref{AppA1} and \ref{AppA2} above, $\omega_{\mbox{\scriptsize max}}^{(1)}$ is characterized by an effective interaction energy $V_{\mbox{\scriptsize eff}}(L)$. It is then natural to ask whether in the finite-range interaction regime, the anharmonicity can also be characterized by $V_{\mbox{\scriptsize eff}}(L)$: here we investigate this question numerically.

\begin{figure}[H]
\centering
\includegraphics[width=0.5\textwidth]{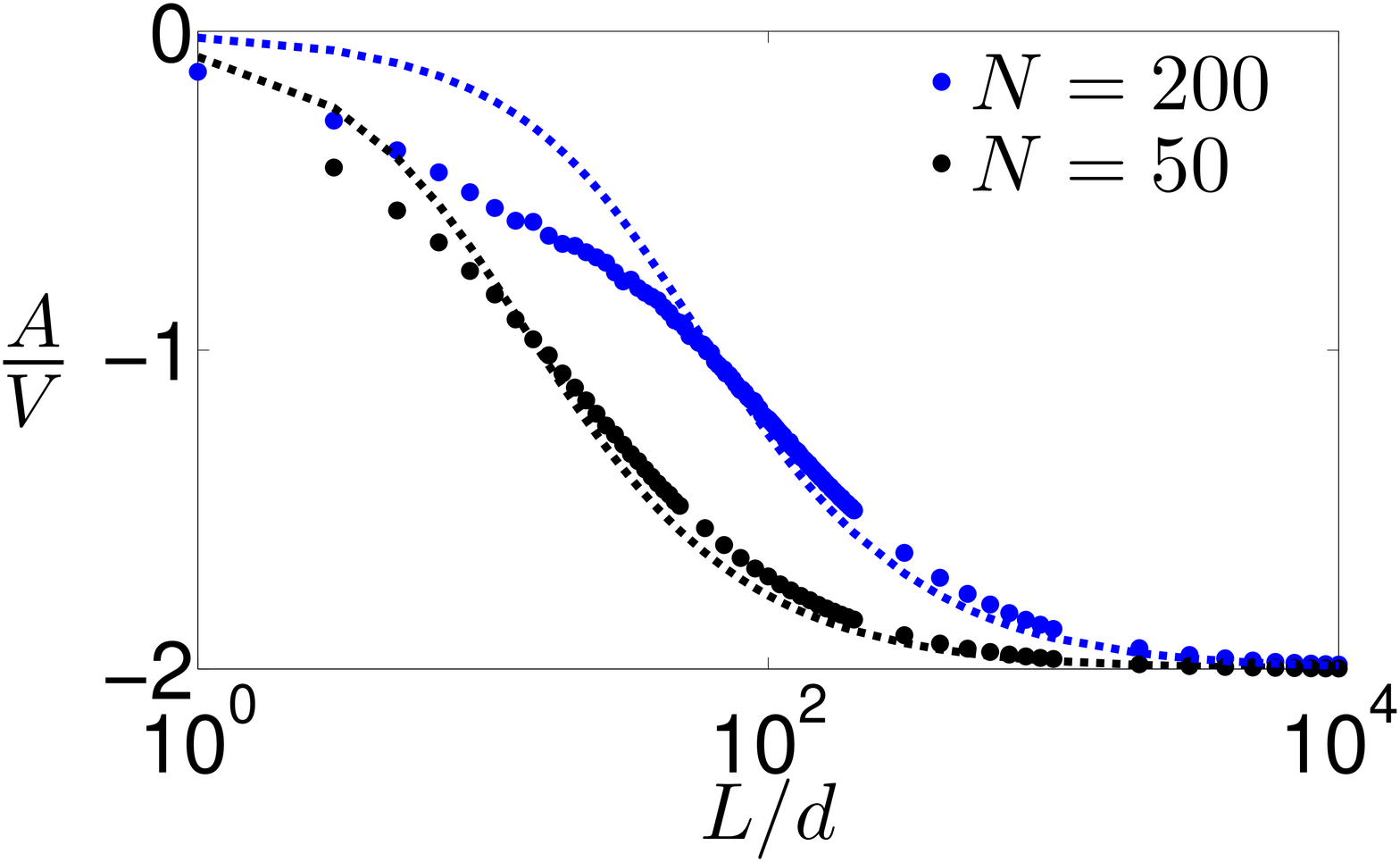}
\caption{Mean anharmonicity $A = \omega_{\mbox{\scriptsize max}}^{(2)}-2\omega_{\mbox{\scriptsize max}}^{(1)}$ from direct diagonalization (1000 samples per $L$), and from the formula $A=-2V_{\mbox{\scriptsize eff}}(L)$ (both normalized to $V$), as a function of $L$ for $N=200$ (blue) and $N=50$ (black), with $n=20$ in both cases.}
\label{anharmonicity}
\end{figure}

Figure \ref{anharmonicity} shows the mean anharmonicity $A$ (solid dots, with averaging over 1000 samples of atomic positions), normalized to the interaction strength $V$, as a function of $L$ for two different system sizes, $N=50$ and $N=200$, with $n=20$ atoms. In addition, we plot (dashed lines) the hypothesized anharmonicity $-2V_{\mbox{\scriptsize eff}}(L)$. In both cases we see that for large $L/d\gtrsim N/2$, the approximation based on $V_{\mbox{\scriptsize eff}}(L)$ is very accurate, while for small $L$ there is a significant difference between the two models, which clearly increases for larger systems.

\subsection{Effective model for many-photon dynamics}\label{AppB3}

Here we describe the effective model for the multi-photon behavior, which may be used to make approximate predictions when either the number of atoms or excitations becomes sufficiently large that solutions of the full master equation become infeasible. We build on the linear model described in Section \ref{LinearToyModel}, assuming now that the system may be described in terms of two bosonic-type modes: (1) a mode corresponding to the `ensemble' of $n$ independent atoms, which can support up to $n$ excitations, and; (2) a spin-wave corresponding to the maximum resonance $|1\rangle$ of the first-excited manifold, and its doubly-excited state $|2\rangle$ in the two-photon manifold. The Hamiltonian is given by

\begin{eqnarray}\label{EffHamAppB}
H_{\mbox{\scriptsize eff}} &=& \sum_{m=1}^n \omega_m \sigma_{E_m,E_m} + \sqrt{n}\Omega\sum_{m=1}^n\sqrt{m}(\sigma_{E_m,E_{m-1}} + h.c.) \\ \nonumber
&+& \Delta_{\mbox{\scriptsize max}}\sigma_{11} + 2(\Delta_{\mbox{\scriptsize max}} + V_{\mbox{\scriptsize eff}})\sigma_{22} + \sqrt{n\kappa_n}\Omega\left(\sigma_{1g} + \sqrt{2}\sigma_{21} + h.c.\right).
\end{eqnarray}

Here, $\sigma_{\mu\nu}= |\mu\rangle\langle \nu|$ for energy eigenstates $|\mu\rangle$ and $|\nu\rangle$, $|E_m\rangle$ denotes the $m^{th}$ excited state of the ensemble, with corresponding energy $\omega_m = -m(\Delta_{\mbox{\scriptsize max}} + nV_{\mbox{\scriptsize eff}})$, and $|E_0\rangle = |g\rangle$, and all other quantities are as defined in the effective model of Appendix \ref{AppA4}. Dissipation is described by the Lindblad operator

\begin{eqnarray}
\mathcal{L}_{\mbox{\scriptsize eff}}[\rho] &=& \frac{\Gamma_n}{2}\sum_{m=1}^{n}m\left(\sigma_{E_m}\sigma_{E_m}\rho + \rho \sigma_{E_m,E_m}\sigma_{E_m,E_m} - 2\sigma_{E_{m-1},E_m}\rho\sigma_{E_m,E_{m-1}}\right) \\ \nonumber
&+& \frac{\gamma_n}{2}\left(\sigma_{11}\rho + \rho\sigma_{11} - 2\sigma_{g1}\rho\sigma_{1g}\right) \\ \nonumber
&+& \gamma_n\left(\sigma_{22}\rho + \rho\sigma_{22} - 2\sigma_{12}\rho\sigma_{21}\right) \\ \nonumber
\end{eqnarray}

where $\Gamma_n = \Gamma^{\prime} + n\Gamma_{\g}$ is the decay rate of the ensemble, and $\gamma_n = \Gamma^{\prime} + n\kappa_n\Gamma_{\g}$ is the decay rate of excitations in the spin wave. We now illustrate the predictive power of the effective model, numerically solving the master equation $\dot{\rho} = -i[H_{\mbox{\scriptsize eff}},\rho] + \mathcal{L}_{\mbox{\scriptsize eff}}[\rho]$ in two example cases, and comparing with the results of the full model of Eq. \ref{Htot} in the main text. Furthermore, in Appendix \ref{AppB4} we use the effective model to obtain simple analytical results in the regime where the system is driven close to the resonance frequency $\omega_{\mbox{\scriptsize max}}$.

\begin{figure}[H]
\centering
\subfigure[]{
\label{maxpop_func_atoms_withtoy}
\includegraphics[width=0.45\textwidth]{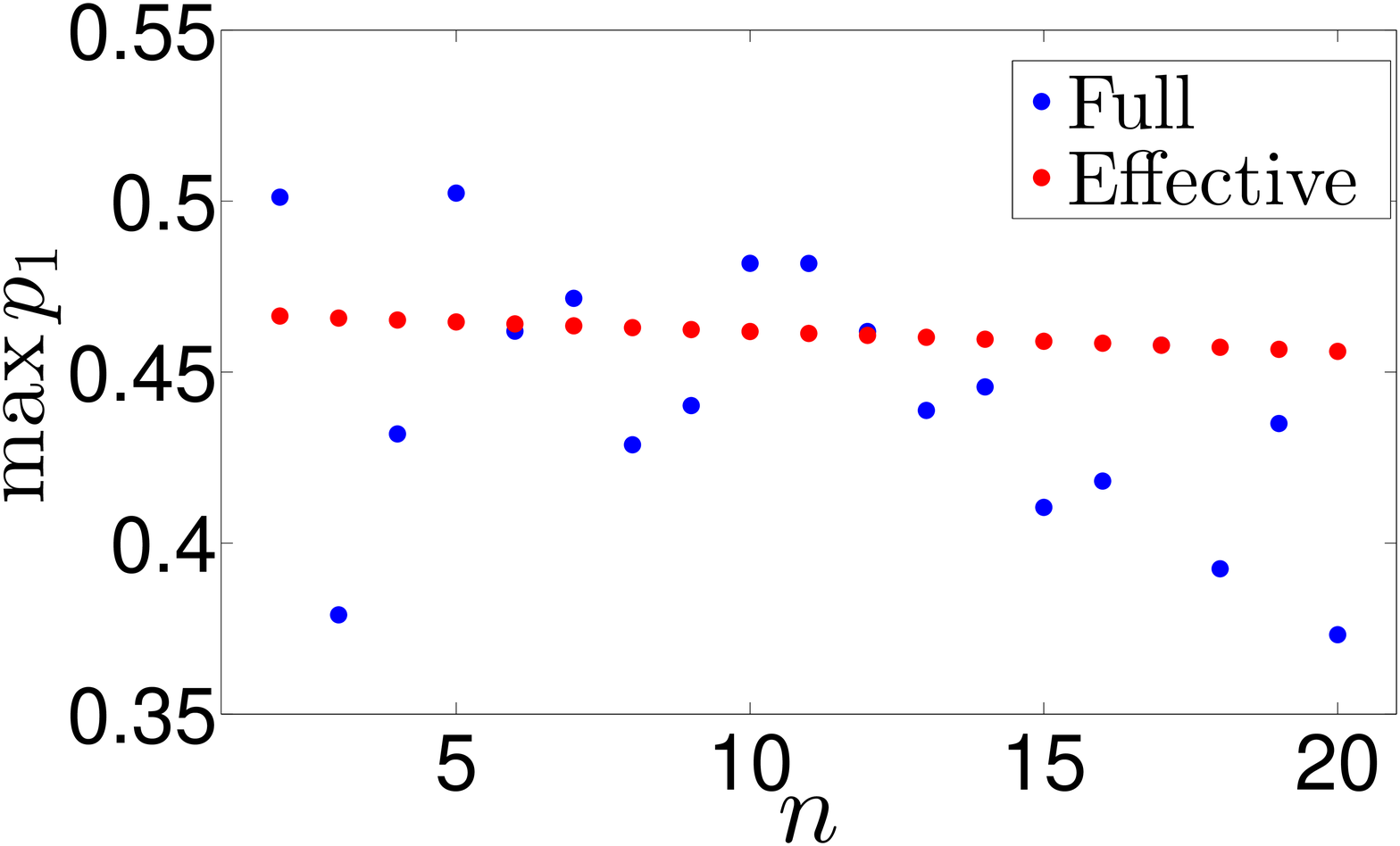}}
\subfigure[]{
\label{OVERLAP_diff_full_effective}
\includegraphics[width=0.45\textwidth]{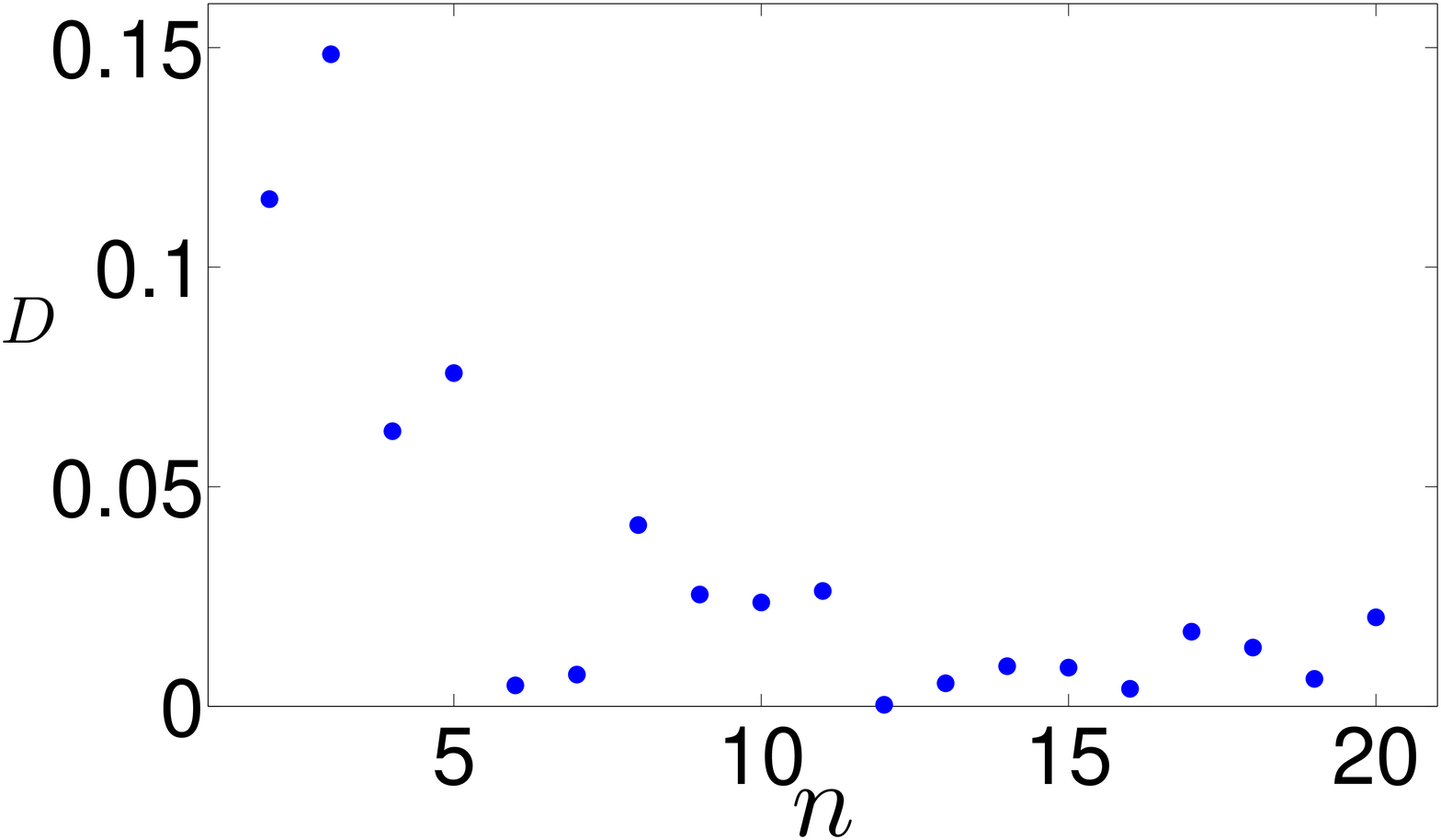}}
\caption{(a) Blue dots: full model prediction (with Hilbert space truncated at the second excited manifold) for the maximum single-excitation fidelity (optimized over the detuning $\Delta_{\mbox{\scriptsize max}}$) as a function of number of atoms, for $\Omega/\Gamma^{\prime}=1$, $N=200$, $L/d=10^6$, $V/\Gamma^{\prime}=10$, $k_ad = \pi/2$, $\Gamma_{\g}/\Gamma^{\prime}=0.3$. Red dots: prediction from effective model for the same parameters. (b) Difference $D$ between the overlap $\sqrt{\kappa_n}$ used in the effective model, and the overlap $\langle\phi_{\mbox{\scriptsize max}}|\psi_{\mbox{\scriptsize in}}\rangle$ for each of the cases in (a).}
\label{ToySpincompare}
\end{figure}

Fig. B\ref{maxpop_func_atoms_withtoy} shows the predictions of both the effective and full models for the maximum first-excited state population as a function of the number of atoms $n$, optimized over the detuning $\Delta_{\mbox{\scriptsize max}}$, for a driving strength $\Omega/\Gamma^{\prime}=1$. Here, in order to obtain the solution for the full model, we first verify that for small $n\lesssim 6$, the dynamics for this particular choice of $\Omega$ are very well approximated by truncating the Hilbert space at the second excitation manifold (i.e. by discarding states with 3 or more excitations). Since for larger values of $n$ the maximum-energy resonance is driven more weakly (due to the reduced overlap with the input field), and hence the probability of introducing additional excitations is suppressed, we then obtain solutions for up to $n=20$ with the truncated Hilbert space.

Rather than averaging the dynamics over many configurations, in this case we simply search through 1 million samples of random atomic positions and, for each $n$, choose the configuration whose maximum single-photon resonance $|\phi_{\mbox{\scriptsize max}}\rangle$ has an overlap with the input excitation $|\psi_{\mbox{\scriptsize in}}\rangle$ that is closest to the overlap $\sqrt{\kappa_n}$ used in the effective model. In other words, for each $n$ we seek the configuration of atoms that minimizes the quantity $D = |\sqrt{\kappa_n} - \langle\phi_{\mbox{\scriptsize max}}|\psi_{\mbox{\scriptsize in}}\rangle|$.

We see from the figure that the two models agree reasonably well in their predictions for the maximum fidelity of introducing only a single excitation, with an average error (i.e. the average absolute difference between the values predicted by the models, normalized to the value from the full model) of approximately 7.5\%. Since in general it is not possible to find a configuration of $n$ atoms such that the difference in overlaps $D$ is identically zero, we expect there to be some degree of discrepancy between the two cases. Figure B\ref{OVERLAP_diff_full_effective} shows $D$ for each of the cases in Fig. B\ref{maxpop_func_atoms_withtoy}, illustrating that the agreement between the two models is best when $D$ is smallest.

\begin{figure}[H]
\centering
\includegraphics[width=0.5\textwidth]{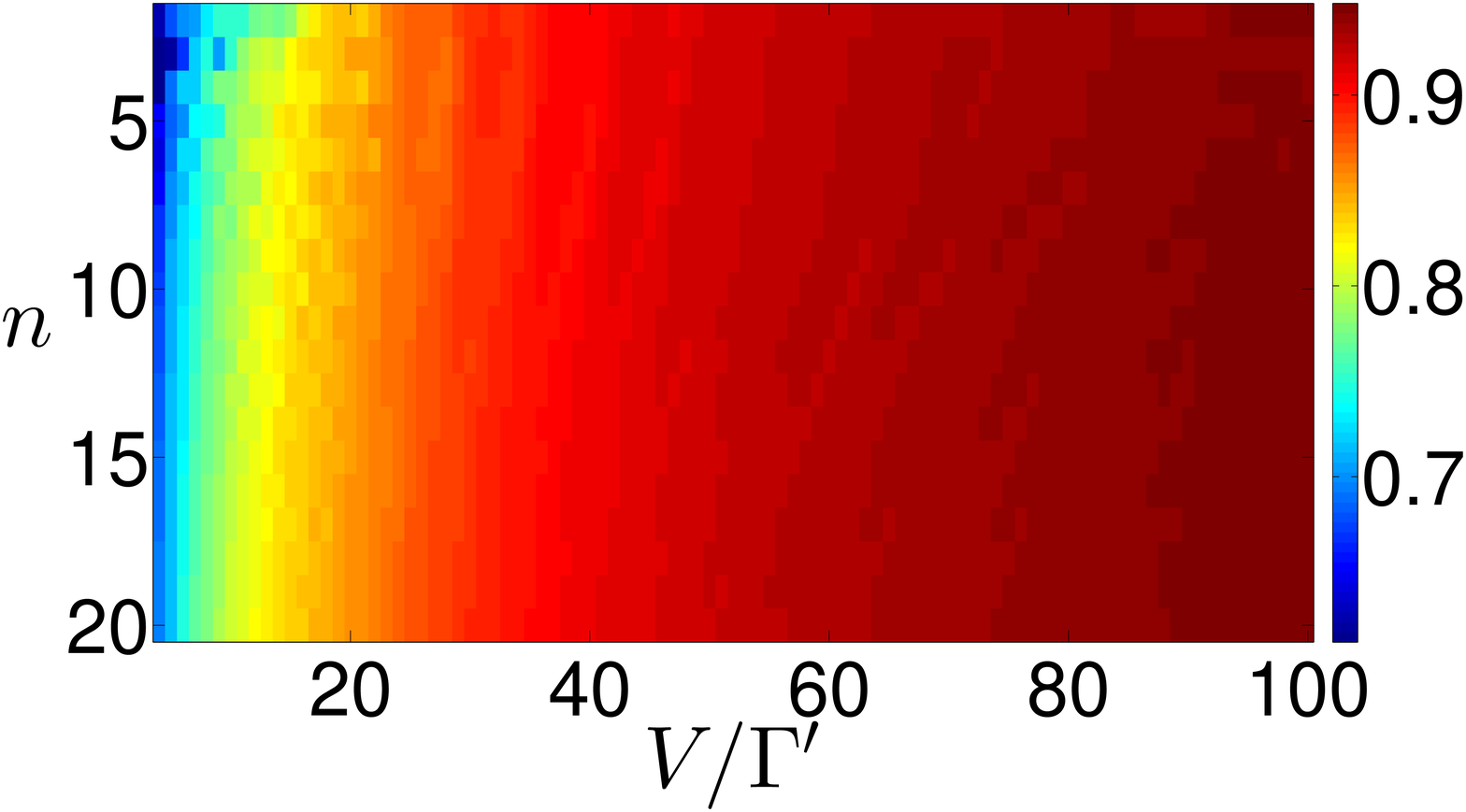}
\caption{Maximum single-excitation fidelity, optimized over Rabi frequency $\Omega$ and detuning $\Delta_{\mbox{\scriptsize max}}$, as a function of interaction strength and number of atoms, obtained using the effective model. Other parameters: $L/d = 10^6$, $N=50$, $\Gamma_{1D}/\Gamma^{\prime} = 0.3$.}
\label{contour_Vandn}
\end{figure}

Having verified the validity of our effective model, we now apply it to the situation described in the main text. In particular, we apply a global Rabi pulse to all atoms at a frequency near the maximum resonance, and consider the fidelity of producing only a single excitation. Figure \ref{contour_Vandn} shows the prediction for the optimal single-excitation probability $p_1$ as a function of interaction strength $V$ and number of atoms $n$, optimized over both driving strength $\Omega$ and detuning $\Delta_{\mbox{\scriptsize max}}$. We see that as the interaction strength increases, the fidelity of introducing a single excitation is very high regardless of the number of atoms, since by increasing the strength of the driving one can overcome the weaker coupling of the input field to the spin wave.

\subsection{Analysis of single-excitation probability for driving close to $\omega_{\smax}$}\label{AppB4}

In this section, we use the effective model to analytically quantify the errors that reduce the fidelity of introducing only a single excitation. We assume that the dominant errors are (1) dissipation via emission into free space and into the waveguide, and (2) population of the doubly-excited manifold. We begin by considering the latter type, and will show that under the appropriate conditions these errors become negligible, in which case the fidelity is then limited only by errors of type (1).

The system can be doubly-excited by acquiring population either in the state $|2\rangle$, or in the second excited state $|E_2\rangle$ of the ensemble. In order to illustrate how the populations of these states scale with the system parameters, we assume the system to be lossless and solve the Schr\"{o}dinger equation (SE) for the atomic wavefunction: the generalization to the case with dissipation is straightforward, and does not qualitatively affect the following conclusions. The SE reads $i\dot{\vec{c}} = H\vec{c}$, where the state vector $\vec{c} = \left(c_g,c_{E_1},c_1,c_{E_2},c_2\right)^T$, $c_i$ are the probability amplitudes of the energy eigenstates (labelled by $i$) of the model, and the Hamiltonian is given by Eq. \ref{EffHamAppB}.

We are interested in the regime where there is a high probability of exciting state $|1\rangle$, so by taking $\Omega\ll\sqrt{n}V_{\mbox{\scriptsize eff}}$ we can strongly suppress excitation of the ensemble. Solving for the amplitudes $c_{E_1}$ and $c_{E_2}$, we find (taking - for simplicity - $L/d\rightarrow\infty$, such that $V_{\mbox{\scriptsize eff}}\rightarrow V$)

\begin{eqnarray}
c_{E_1} &=& \frac{\sqrt{n}\Omega}{nV + \Delta_{\mbox{\scriptsize max}}}\left(1 - \frac{n\Omega^2}{2(nV+\Delta_{\mbox{\scriptsize max}})^2}\right)^{-1}c_g \\ \nonumber
c_{E_2} &=& \frac{n\Omega^2}{2(nV + \Delta_{\mbox{\scriptsize max}})^2}\left(1 - \frac{n\Omega^2}{2(nV+\Delta_{\mbox{\scriptsize max}})^2}\right)^{-1}c_g \\ \nonumber
\end{eqnarray}

Inserting the solution for $c_{E_1}$ into the equation of motion for $c_g$, to lowest order in $\Omega/nV$ we find

\begin{equation}
\dot{c_g} \sim -i \frac{n\Omega^2}{nV + \Delta_{\mbox{\scriptsize max}}}c_g -i\Omega_n c_1
\end{equation}

where we define $\Omega_n = \sqrt{n\kappa_n}\Omega$. The first term on the RHS is a Stark shift of the ground state energy due to the off-resonant driving of the ensemble, while the second term describes excitation to the state $|1\rangle$ with Rabi frequency $\Omega_n$. An appropriate choice of $\Delta_{\mbox{\scriptsize max}}$ (with $\Delta_{\mbox{\scriptsize max}} >0$) may be used to ensure the latter process dominates.

Turning to the effect of coupling to state $|2\rangle$, from the SE we have

\begin{equation}
\dot{c_2} = -i\Omega_n c_1 - 2i(V + \Delta_{\mbox{\scriptsize max}})c_2
\end{equation}

Again, by choosing $\Delta_{\mbox{\scriptsize max}}$ appropriately, with $\Delta_{\mbox{\scriptsize max}} >0$, the Rabi oscillations between states $|1\rangle$ and $|2\rangle$ may be neglected, allowing for adiabatic elimination of the variable $c_2$, giving $c_2 = -\Omega_nc_1/2(V + \Delta_{\mbox{\scriptsize max}})$. As a result, we find

\begin{equation}
\dot{c_g} = -i\Omega_n c_1 \qquad \dot{c_1} = -i\Omega_n c_g +i\frac{\Omega_n^2}{2(V+ \Delta_{\mbox{\scriptsize max}})}c_1
\end{equation}

In the second equation here, the state $|1\rangle$ is Stark-shifted as a result of the off-resonant drive of state $|2\rangle$. In order for this to be negligible compared to the Rabi oscillations between the states $|1\rangle$ and $|g\rangle$, we require $\Omega_n/2(V + \Delta_{\mbox{\scriptsize max}})\ll1$, which may again be satisfied by choosing the detuning appropriately.

The conclusion of this analysis is therefore that the population $p_1 = |c_1|^2$ of state $|1\rangle$ undergoes Rabi oscillations, and is followed by the population $p_2 = |c_2|^2$ of state $|2\rangle$, with a magnitude that is suppressed with respect to $p_1$ by an amount $\sim\Omega_n^2/4V^2$. Meanwhile, errors due to population of the ensemble are suppressed by a factor of $\sim \Omega^2/nV^2$ at the single-photon level, and by a factor of $\sim \Omega^4/4n^2V^4$ at the two-photon level.

\begin{figure}[H]
\centering
\includegraphics[width=0.6\textwidth]{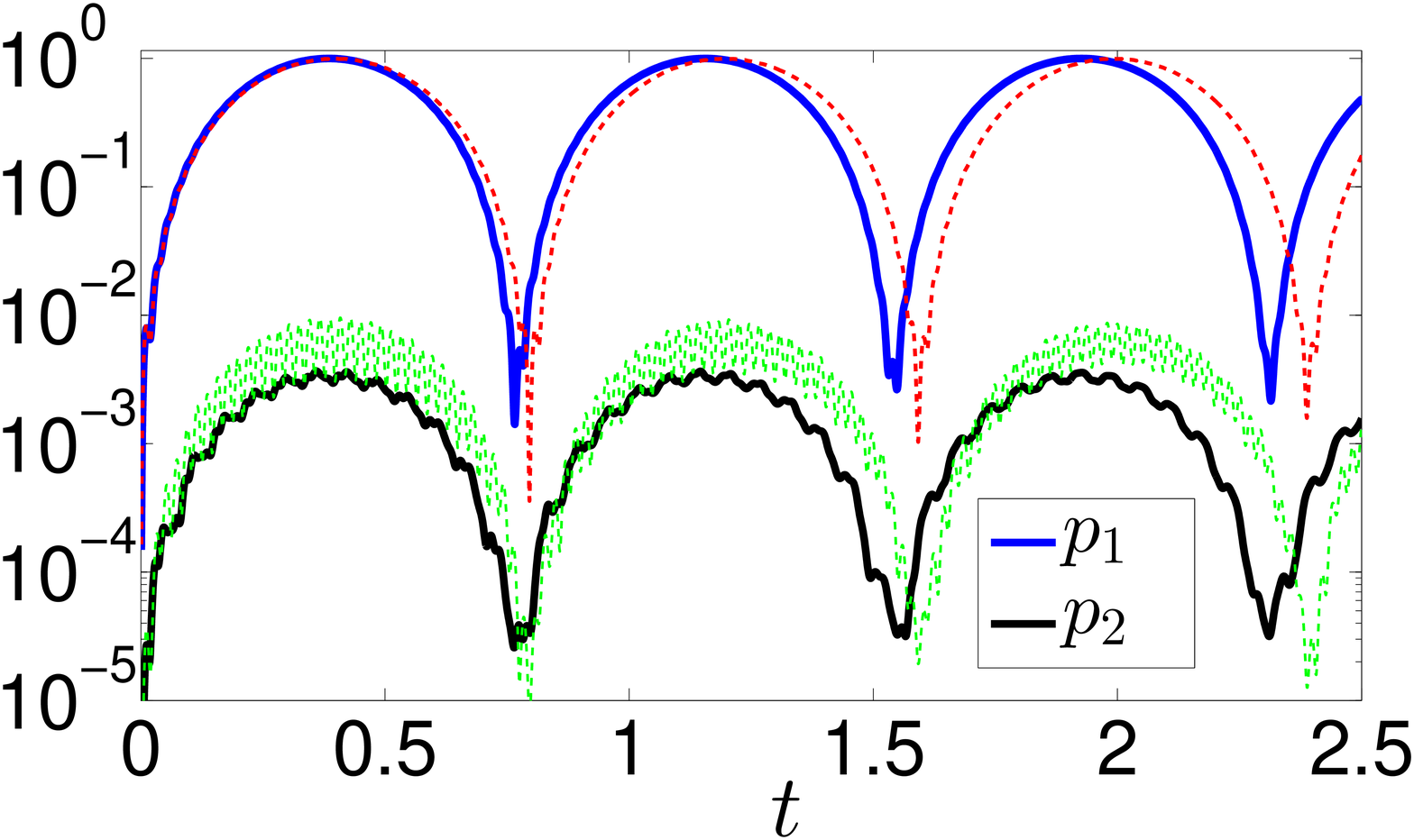}
\caption{Solid lines: time evolution of the populations of the first (blue) and second (black) excited manifolds in the full model, for the lossless case $\Gamma_{\g} = \Gamma^{\prime}=0$. Dashed curves: effective model predictions (red for first excited manifold, green for second). Parameters used: $\Omega=5$, $\Delta_{\smax}=0$, $N=50$, $L/d = 10^6$, $V=50$, $n=6$, with the positions chosen such that the overlap $\langle\phi_{\mbox{\scriptsize max}}|\psi_{\mbox{\scriptsize in}}\rangle\approx \sqrt{\kappa_6}$.}
\label{population_levels1and2}
\end{figure}

To illustrate this, in Fig. \ref{population_levels1and2} we plot the full model solution (with $\Gamma^{\prime}=\Gamma_{\mbox{\scriptsize 1D}} = 0$) for the total populations in the first ($p_1$) and second ($p_2$) excited manifolds (taking $N=50$, $V/\Omega = 10$, and $n=6$, using a configuration with overlap $\langle\phi_{\mbox{\scriptsize max}}|\psi_{\mbox{\scriptsize in}}\rangle\approx \sqrt{\kappa_6})$, as well as the solutions from the effective model under the same conditions. We see that the models agree well, and moreover when $p_1\sim1$, the population $p_2\sim 1/300$, which is of the order the value we would expect based on the analytically derived ratio $p_2/p_1 \sim \Omega_n^2/4V^2 \approx 1/600$.

In this regime, where the ensemble behaves as an effective two-level atom with states $|g\rangle$ and $|1\rangle$, the dominant errors are due to the decay of the spin wave $|1\rangle$ into free space and the waveguide, i.e. errors of type (1), as defined above. We can quantify this error by numerically calculating the maximum population that can be inverted in an ideal two-level system with the same detuning, drive strength, and decay rate. In other words, the error in achieving full inversion may be found by numerically solving the simple two-level system master equation $\dot{\rho} = -i[H,\rho] + \mathcal{L}[\rho]$, where

\begin{eqnarray}
H &=& \Delta_{\mbox{\scriptsize max}}\sigma_{11} + \Omega_n\left(\sigma_{g1} + \sigma_{1g}\right) \\ \nonumber
\mathcal{L}[\rho] &=& \frac{\gamma_n}{2}\left(\sigma_{11}\rho + \rho\sigma_{11} - 2\sigma_{g1}\rho\sigma_{1g}\right)
\end{eqnarray}

\end{document}